\documentclass[12pt]{article}
\usepackage{amsfonts}
\usepackage{amssymb}
\usepackage{epsfig}
\usepackage{dcolumn}
\usepackage{color}
\makeatletter

 \@addtoreset{equation}{section}
\makeatother

\usepackage{amsmath}
\numberwithin{equation}{section}

\advance \textheight by \topskip
\usepackage{amsmath,dsfont}
\usepackage[english]{babel}

\usepackage[colorlinks]{hyperref}
\hypersetup{
    colorlinks=true,
    linkcolor=blue,
    filecolor=magenta,  
    urlcolor=[rgb]{0.00,0.00,0.50},    
    citecolor=red,
    }

\usepackage{url}

\def\C{\mathbb C}
\def\R{\mathbb R}

\def\1{\mathbb 1}

\def\be{\begin{equation}}
\def\ee{\end{equation}}

\begin{document}

\title{
{\bf 
Nonlinear 
symmetries of perfectly 
invisible $PT$-regularized conformal and superconformal 
mechanics systems
 }}

\author{{\bf Juan Mateos Guilarte${}^a$ and Mikhail S. Plyushchay${}^b$} 
 \\
[8pt]
{\small \textit{
${}^a$Departamento de F\'{\i}sica Fundamental and IUFFyM, Universidad de Salamanca,}}\\
{\small \textit{ Salamanca E-37008, Spain}}\\
{\small \textit{ ${}^b$Departamento de F\'{\i}sica,
Universidad de Santiago de Chile, Casilla 307, Santiago,
Chile  }}\\
[4pt]
 \sl{\small{E-mails:  \textcolor{blue}{guilarte@usal.es}, 
\textcolor{blue}{mikhail.plyushchay@usach.cl}
}}
}
\date{}
\maketitle

\begin{abstract}
We  investigate how 
the Lax-Novikov integral in the perfectly invisible $PT$-regularized zero-gap 
quantum conformal and superconformal mechanics systems  affects  on  their  
(super)-conformal symmetries. 
We show that the expansion of the conformal symmetry with this integral 
results in  a nonlinearly  extended generalized Shr\"odinger algebra.
The $PT$-regularized superconformal mechanics systems
in the phase of the unbroken exotic nonlinear $\mathcal{N}=4$ super-Poincar\'e symmetry
are described by nonlinearly super-extended Schr\"odinger  algebra
with the $osp(2|2)$  sub-superalgebra.
In the partially broken phase, the scaling dimension
of  all odd integrals is indefinite, and the
$osp(2|2)$  is not contained as a sub-superalgebra. 
\end{abstract}


\section{Introduction and summary}\label{Intro}

Conformal mechanics model 
was introduced and investigated  
by de Alfaro, Fubini and Furlan  (AFF) \cite{deAFF} as 
a (0+1)-dimensional conformal field theory.  It corresponds  
to the  two-particle  Calogero  system \cite{Calogero}  with eliminated 
center of mass degree of freedom. 
Supersymmetric extension of the model was considered
in  \cite{AkuPash} and  
\cite{FubRabi}.
The geometric aspects of conformal and superconformal 
mechanics were investigated  
in \cite{IvaKriLev1,IvaKriLev2}.
The many-particle generalizations of superconformal mechanics 
were studied 
in \cite{FreeMen,Wyll,BGIK,BGK,BGL,GLP,KL}. 
A revival  of interest to (super)conformal 
mechanics was induced in connection with the AdS/CFT correspondence
\cite{AdS/CFT1,AdS/CFT2,AdS/CFT3} when it was observed that 
the dynamics of a superparticle near the horizon of an
extreme Reissner-Nordstr\"om black hole 
is described by superconformal mechanics \cite{CDKKTV,AIPT,GibTow,MichStro}.
In the same line of the AdS/CFT correspondence,
recently superconformal 
mechanics was employed in the study of physics underlying 
the confinement dynamics in QCD \cite{deTerDosBro,BTDL}.
For some further references and reviews  on conformal and superconformal mechanics
see  \cite{DHVin}--\cite{InzPly}. 
\vskip0.05cm

The conformal mechanics model without confining potential term,
\be\label{Hg}
H_g=-\frac{d^2}{dx^2}+\frac{g}{x^2}\,, \qquad
x>0\,,
\ee
is related  at special values of the coupling constant  $g=n(n+1)$, $n=1,2,\ldots$, 
to the  Korteweg-de Vries (KdV) equation\,:
at $n=1$ its potential is a solution of the stationary KdV equation,
while for $n>1$ it satisfies the $n$-th stationary equation of the 
KdV hierarchy.
This is related to  a broader picture 
according to which the Calogero-Moser systems govern the dynamics of the moving
poles of rational solutions to the KdV equation \cite{AirMcKMos,AdlMos,GorNek}.
The conformal mechanics (\ref{Hg}) with coupling constant $g=n(n+1)$ 
plays also a special role
in the bispectral problem \cite{DuiGru}  as well as in  the 
Huygens'  principle \cite{Veselov}.
\vskip0.05cm

The potential of conformal mechanics model with the indicated
special values of the coupling constant can be obtained 
from  the potentials of the quantum reflectionless  and finite-gap   systems
via appropriate  complex shift of the argument and subsequent  application of  a
certain limit procedure \cite{MatPly}.
One-dimensional quantum reflectionless  and finite-gap   systems, in turn,  
are  the algebro-geometric  solutions to equations of the KdV hierarchy  via the Lax
pair representation \cite{NovZak,Krich,BelBob}. 
Each such   quantum system is characterized by a nontrivial Lax-Novikov integral
of motion which is a  differential operator of  odd order
$2n+1\geq 3$. There exists no
classical analog for this integral having  a purely  quantum origin and nature.
It detects all the non-degenerate bound and edge-states 
by annihilating them,
and distinguishes  the  doubly degenerate states
inside the valence and conduction/scattering  bands in the spectrum
of a corresponding quantum system. 
In the case of reflectionless systems 
the integral admits a representation   in the form of
a Darboux-dressed momentum operator of the free quantum 
particle~\cite{CorJakPly,CPLax,CNPLax}.
\vskip0.05cm

The Lax-Novikov integral  also plays 
an important role in supersymmetric constructions. 
According to the Burchnall-Chaundy
theorem \cite{BC,Ince}, the square of the Lax-Novikov integral of differential order
$2n+1$ is equal to a (spectral) polynomial of the same order $2n+1$
in Hamiltonian operator of the corresponding quantum 
system. As a consequence,  each non-extended (``purely bosonic") 
 finite-gap or reflectionless quantum mechanical system is characterized by a hidden 
 nonlinear bosonized supersymmetry \cite{CPLax,CNPLax}.
On the other hand, it is because of this integral that
the convensional $\mathcal{N}=2$ supersymmetry of
the pairs of the Darboux-intertwined  reflectionless or finite-gap   
quantum systems  expands  up to an exotic nonlinear $\mathcal{N}=4$ 
supersymmetric structure 
which includes the matrix Lax-Novikov integral  as a bosonic central charge
\cite{CorJakNP,AraMatPly}.
\vskip0.05cm

A rather natural  
question  is therefore if  
the Lax-Novikov integral can be identified for the conformal 
mechanics model with special values of the coupling constant, and if so,
how such integral  could influence on the conformal and superconformal 
symmetries.
\vskip0.05cm

Some time ago it was observed   that the conformal mechanics model with 
coupling constant  $g=n(n+1)$ possesses a differential operator $P_n$
of order $2n+1$, which commutes with the Hamitlonian $H_{{g}}$
and satisfies a Burchnall-Chaundy  relation of the form 
$(P_n)^2=(H_{{g}})^{2n+1}$ 
\cite{LeiPly}.
 There appears an obstacle, however,  that the  operator 
$P_n$ is \emph{not physical} from the point of view of quantum mechanics\,: 
as it was shown in \cite{CorOlPly}, 
acting on eigenstates of the 
conformal mechanics model satisfying the Dirichlet boundary 
condition at $x=0$, it transforms them into
formal,  non-physical eigenstates of the Hamiltonian operator
which satisfy Neumann boundary condition at $x=0$.

In spite of a non-physical nature 
of the operator $P_n$ from the point of view of 
the quantum mechanical system $H_g$ with $g=n(n+1)$,
it is the Lax-Novikov operator which underlies the  
above-mentioned relation of conformal mechanics 
to the KdV equation and its hierarchy.
The Burchnall-Chaundy  relation
means an algebraic dependence between 
$H_g$ and the formal  integral $P_n$,
and so, the presence of the latter does not influence 
on integrablity of one-dimensional quantum systems (\ref{Hg}),
which, as any other one-dimensional quantum system with conserved
Hamiltonian, is (maximally) super-integrable.
However, this relation has a super-algebraic nature
implying  that the Lax-Novikov operator is a kind of a square root
operator  from the odd order polynomial (monomial here)
in Hamiltonian.

In a recent paper \cite{MatPly}
it was shown that the indicated deficiency 
of the Lax-Novikov operator $P_n$ can be cured by
the $PT$-regularization of the conformal mechanics model 
by making a shift $x\rightarrow x+i\alpha$, 
where 
$\alpha$ is a nonzero real parameter, and extending 
$x$ from the half-line $x>0$  
to the whole real line $x\in\R$.
With such a shift and extended domain, the Hamiltonian  
$H_n(x+i\alpha)=-\frac{d^2}{dx^2}+\frac{n(n+1)}{(x+i\alpha)^2}$ 
is  $PT$-symmetric 
satisfying  the relation $[PT, H_n(x+i\alpha)]=0$ \cite{BenRev,Most}.
Here $P$  is a space reflection  (parity) operator, $P x = -xP,$ $P^2=1$,
and a complex conjugation operator $T$ is defined by $Tz = \bar{z}T$,
 $T^2 = 1$, where $z\in \C$ is an arbitrary
complex number~\footnote{The peculiarity of $PT$-symmetry 
is associated with the anti-unitary character of the 
operator $T$ \cite{Wigner}.}.
The  obtained quantum systems $H_n(x+i\alpha)$ 
are characterized by the property  of the perfect invisibility\,: 
the transmission amplitude in them  is not  simply just a phase
as it happens in the 
case of reflectionless systems, but exactly equals one like 
in the free quantum particle system.
Unlike the free quantum particle, however, 
 each  such a system has a unique 
quadratically integrable bound   state of zero energy 
at the very edge of the continuous 
part of the spectrum~\footnote{A  similar picture appears
in  finite-gap systems where
non-periodic defects can produce bound states
at the very edge of the valence and conduction bands \cite{Defects1,Defects2}.}. 
The corresponding systems 
are identified as \emph{perfectly invisible zero-gap} systems,
and they possess some other  interesting properties 
due to their relation to the KdV hierarchy \cite{MatPly}.
It may be noted that the parameter $\alpha$ is a constant with dimension of length, 
and in this aspect the $PT$-regularization of the conformal
mechanics model turns out to be alternative  in some sense  
to regularization  of the conformal
mechanics model considered in the original article \cite{deAFF},
where it  was realized effectively via the introduction into the Hamiltonian
(\ref{Hg})
of the confining harmonic oscillator potential term accompanied by a  
length parameter.

It was  observed for the first time  by
 Bender and Boettcher in the pioneering work  \cite{BenBoe} 
 that Hermiticity   is not a necessary condition for the reality of the spectrum 
of a quantum mechanical system,  and that it can 
be substituted  for the requirement of the $PT$-symmetry of 
Hamiltonian~\footnote{Reality of spectrum in some quantum mechanical systems 
with non-Hermitian Hamiltonian operators was observed 
earlier, but the  reason of this phenomenon associated 
with the presence of $PT$-symmetry was established 
for the first time in  \cite{BenBoe},
see the corresponding discussion in \cite{BenRev}.}.
Later this type of systems was investigated in different aspects,
in particular, in the context of connection between the theories of ordinary
differential equations and integrable models \cite{DDT1},
and supersymmetry \cite{DDT2}.
The  results on reality of the spectrum 
were extended then for non-Hermitian Hamiltonians
of a more general form, for reviews see refs.
\cite{BenRev} and \cite{Most}.
The $PT$-regularization applied in  \cite{MatPly} to 
(\ref{Hg}) 
 was employed earlier    in general context of the $PT$-symmetry
for  the AFF model with confining potential term   \cite{Znoj}.
$PT$-symmetric multi-soliton solutions to the Korteweg-de Vries
equation were discussed recently in
\cite{CorFri},
and $PT$-symmetric deformations of Calogero 
models were studied in \cite{FriZno,CorLech}.
Nowadays $PT$-symmetry
finds interesting applications 
in diverse areas of physics
\cite{NatPhys}. 

\begin{itemize}
\item The purpose of this article is to investigate how 
the Lax-Novikov integral in the $PT$-regularized quantum conformal
and superconformal mechanics models  affects  on  their  
(super)-conformal symmetries. 
\end{itemize}

It is necessary to stress here  that the  $PT$-regularization 
$x\rightarrow x+i\alpha$  cannot be considered 
as a kind of a simple analytic continuation of the models (\ref{Hg})
since the domain of the latter  corresponds to  the half-line, while the $PT$-regularized
systems are defined on the whole real line $x\in\R$.
As we shall see, this  results in essential difference in symmetry properties 
of the $PT$-regularized systems $H_n(x+i\alpha)$
and of their supersymmetric versions in comparison with (\ref{Hg}) and its
superextensions. The essential difference can also be expected {\it a priori}
if, by analogy, we  compare the properties of the quantum systems described by 
potentials  $u_1=\frac{n(n+1)}{\sinh ^2 x}$ and $u_2=-\frac{n(n+1)}{\cosh^2 x}$.
Both potentials (shifted for appropriate additive constant terms)
are solutions to the corresponding stationary equations of the KdV hierarchy,
and the second quantum system is related to the first one by 
a simple complex shift $x\rightarrow x+i\pi/2$. The first quantum system, however, 
is singular at $x=0$ being defined on the real half-line $x>0$ (or $x<0$),
its non-degenerate spectrum is continuous  and has no ground state.
The second system is reflectionless  being defined on  the whole 
real line, its continuous spectrum is doubly degenerate  (except of the non-degenerate 
state with $E=0$ at the lower edge), and has $n$ non-degenerate negative energies
corresponding to bound states,  with energy of 
the ground state $\psi_0(x)=1/\cosh^n x$ equal to $E_0=-n^2$ \cite{CPLax,CorJakPly}.
\vskip0.1cm

Our results  can be  summarized briefly as follows.
We first show that   the extension  of the set of generators of conformal 
symmetry of the $PT$-regularized  conformal mechanics model with 
coupling constant $g=2$  ($n=1$)
by  its Lax-Novikov integral
generates three more dynamical (explicitly depending on time)
 integrals of motion together with a central charge.
 As a result,   the conformal $so(2,1)\simeq sl(2,\R)$ Lie algebra expands  up to a
nonlinearly extended  Schr\"odinger  algebra, in which the $so(2,1)$ generators
appear quadratically in commutation relations of the four new 
non-trivial (including Lax-Novikov)  integrals of motion. 
With respect to the adjoint action of the $so(2,1)$ generators, the complete set of
integrals separates into 
one-dimensional representation corresponding to
the mass central charge, while the rest of generators are eigenstates
of eigenvalues $(-3/2,-1,-1/2,0,1/2,1,3/2)$ 
of the dilatation generator.
The simplest  supersymmetric  extension of the 
$PT$-regularized  $g=2$  conformal mechanics model is
realized via a usual  construction of 
 $\mathcal{N}=2$ supersymmetric quantum mechanics
based on   superpotential 
$\mathcal{W}_e=-1/(x+i\alpha)$. 
In this case the $PT$-regularized conformal mechanics model 
is paired with the free particle, and the system is described by
the exotic nonlinear $\mathcal{N}=4$ super-Poncar\'e algebra 
which corresponds to  the phase of \emph{exact}, 
unbroken supersymmetry with a non-degenerate 
zero energy ground 
state. The matrix Lax-Novikov integral is 
generated by anticommutator  of the 
first and second order supercharges. 
The extension of the set of generators of 
the exotic nonlinear $\mathcal{N}=4$ super-Poncar\'e 
algebra by the matrix generators of the dilatations and special conformal 
symmetry  transformations 
gives rise to the nonlinearly extended generalized super-Schr\"odinger 
algebra. 
The set of its nontrivial bosonic generators includes the
matrix generalization of the above mentioned seven integrals of motion,
the generator of a $u(1)$ $R$-symmetry, and  two more integrals 
which are the momentum and the Galileo boost generator
of the free particle subsystem. 
The set of the fermionic generators includes three pairs of 
dynamical integrals of motion in addition to the two pairs 
of the supercharges of the nonlinear $\mathcal{N}=4$ super-Poncar\'e 
sub-superalgebra. 
The 
superconformal $osp(2\vert 2)$ symmetry 
is the Lie sub-superalgebra, whose 
expansion 
by
 any other even or odd integral of motion results
 in generation of  the whole nonlinearly super-extended Schr\"odinger 
algebra. 
All the even and odd integrals 
form a supermultiplet with respect to the adjoint action
of the generators of the $osp(2\vert 2)$ superconformal  symmetry,
and 
the structure coefficients in the (anti)commutation relations 
between the rest of  the even and odd integrals 
are linear in generators of the conformal
$so(2,1)$ symmetry.
\vskip0.05cm

A general case of the $PT$-regularized conformal mechanics with 
coupling constant $g=n(n+1)$ is characterized  by a symmetry
described by a nonlinearly extended  Schr\"odinger algebra
generated by  $2n+5$ integrals of motion 
whose scaling dimensions 
are $(-(n+1/2),-(n-1/2),\ldots, n-1/2,n+1/2)$, plus a central charge.
The  $\mathcal{N}=2$ supersymmetric  extension of this
$PT$-regularized  conformal mechanics model is
realized by the construction of 
 supersymmetric quantum mechanical system 
given by a superpotential 
$\mathcal{W}_e=-n/(x+i\alpha)$.
The obtained in such a way $2\times 2$ matrix system 
is characterized by the exotic nonlinear $\mathcal{N}=4$
super-Poincar\'e type symmetry, in which the anticommutator 
of the supercharges, being operators of differential orders
$1$ and $2n$, generates the matrix Lax-Novikov integral.
Extension of this nonlinear $\mathcal{N}=4$ super-Poincar\'e algebra by 
generators of dilatations and special conformal  transformations 
results in expansion of  the superalgebra up to
a nonlinear super-extended Shr\"odinger symmetry, which
contains the superconformal $osp(2\vert 2)$ algebra as a Lie sub-superalgebra.
With respect to the adjoint action of the $osp(2\vert 2)$ generators, the rest 
of the $4n+2$ bosonic and  $4n+2$ fermionic integrals
form an irreducible represenation. 
The structure coefficients 
in (anti)-commutation relations between additional 
even and odd  generators are polynomials of order $2n-1$  in generators 
of the conformal $so(2,1)$ symmetry of the system.
\vskip0.05cm

We also consider supersymmetric system  given by the superpotential 
$\mathcal{W}_b=1/(x+i\alpha_1)-1/(x+i\alpha_2)+i/(\alpha_1-\alpha_2)$, 
$\R\ni \alpha_j\neq 0$, $j=1,2$, $\alpha_1\neq \alpha_2$.
It represents a matrix 
system of the Darboux-paired 
$PT$-regularized conformal mechanics models with $g=2$
characterized by different values of the 
shift  parameters $\alpha_1$ and $\alpha_2$.
This system  is described by the spontaneously  \emph{partially 
broken} 
phase of the exotic nonlinear $\mathcal{N}=4$ super-Poncar\'e symmetry. 
The essential peculiarity of the system is that 
all its fermionic generators commute nontrivially 
with the matrix dilatation generator, 
neither  of them has a definite scaling dimension.
The superalgebra of the system represents some nonlinear 
super-extension of the Schr\"odinger symmetry
which\emph{ does not} include the superconformal $osp(2|2)$ symmetry 
as a sub-superalgebra.
Its generators transform nontrivially  into those 
of the system  given by  superpotential 
$\mathcal{W}_e=-1/(x+i\alpha)$ in the limit when one of the $PT$-regularization
 parameters
is sent  to infinity.  
\vskip0.05cm

The paper is organized as follows.
In Section \ref{Section2}, we describe briefly the 
construction of the perfectly invisible  $PT$-regularized zero-gap
conformal mechanics systems together with their 
Lax-Novikov integrals. 
The nonlinearly extended  Schr\"odinger symmetry of
the $PT$-regularized conformal mechanics model with 
coupling constant  $g=2$ is discussed
in Section \ref{SecSymH1}.
Symmetries of  the simplest superconformal extension of this 
system are investigated 
in Section \ref{SecSusyExact}. 
The results of the two previous sections are generalized for the case 
of $g=n(n+1)$ in Section \ref{Section5}.
The case of the super-extended  conformal mechanics with $g=2$
in the phase of spontaneously partially broken
exotic nonlinear $\mathcal{N}=4$ super-Poincar\'e symmetry 
is considered in Section 
\ref{spontbreake}. 
The concluding discussion is presented in Section  \ref{Outlook}.

\section{$PT$-regularized conformal mechanics models }\label{Section2}

Let us re-denote  the Hamiltonian operator of the conformal mechanics model 
(\ref{Hg})
with non-negative coupling constant  $g=\nu(\nu+1)$,
$\nu\geq 0$, as $H_\nu$.
It admits  two factorizations $H_\nu=A_\nu A_\nu^\#=A_{\nu+1}^\# A_{\nu+1}$
in terms of the first order differential operators $A_\nu=x^\nu\frac{d}{dx}x^{-\nu}=\frac{d}{dx}-\frac{\nu}{x}$, 
$A_\nu^\#=A_\nu^\dagger=-\frac{d}{dx}-\frac{\nu}{x}$.  
The $A_\nu$ and  $A_\nu^\#$  intertwine the Hamiltonian operators
with different in one  values of the index\,: $A_\nu H_{\nu-1}=H_\nu A_\nu$,
$A_\nu^\# H_{\nu}=H_{\nu-1} A_\nu^\#$.
The system $H_\nu$  is defined on the half-axis $x>0$, 
and the case with $\nu=0$ can be  considered as a limit case $\nu\rightarrow 0$ 
corresponding to the quantum free particle on the half-line
$x>0$ subject to the Dirichlet boundary condition $\psi(0)=0$ 
\cite{MatPly}.
We denote the Hamiltonian operator of such a limit system by $H^+_0$.
The case of integer values of the parameter $\nu=n$ is special in the sense 
that the conformal mechanics model Hamiltonian $H_n$ 
can be intertwined with that of  $H^+_0$,
$\mathcal{A}_n^\# H_n=H^+_0\mathcal{A}_n^\#$, 
$\mathcal{A}_n H_0^+=H_n\mathcal{A}_n$, 
by the 
generators of the Darboux-Crum transformation given by
the $n$-th order differential operators   
$\mathcal{A}_n=A_nA_{n-1}\ldots A_1$  and $\mathcal{A}_n^\#=\mathcal{A}_n^\dagger$.
In correspondence with this,
the eigenstates  $\psi_k^{(0)}(x)=\sin kx$ of $H_0^\dagger$ are mapped into eigenstates
of $H_n$, $\psi^{(n)}_k(x)=\mathcal{A}_n\psi^{(0)}_k(x)$,
$H_n\psi^{(n)}_k=k^2\psi^{(n)}_k$.
Unlike the quantum free particle system $H_0=-\frac{d^2}{dx^2}$, $x\in\R$,
the half-free particle system $H^+_0$ is not translation invariant and 
$P_0=-i\frac{d}{dx}$ \emph{is not} its physical 
operator (observable) since
acting on the states  $\psi_k^{(0)}(x)$ it transforms them
into the wave functions satisfying  the Neumann boundary condition $\psi'(0)=0$ 
instead of the Dirichlet boundary condition.
As a result, instead of the algebra of Schr\"odinger symmetry of the 
free particle on a whole line given by the Hamiltonian $H_0$,
the conformal mechanics systems $H_n$, including the case $H_0^+$, are described
only by the algebra $sl(2,\R)$ of its conformal symmetry subgroup.

By virtue of  the  intertwining relations between $H_n$ and $H^+_0$,
each of the systems $H_n$ possesses a {\it formal}
 integral of motion 
$P_n=\mathcal{A}_n P_0\mathcal{A}_n^\#$, $[P_n,H_n]=0$,
which, similarly to the operator $P_0$   for $H^+_0$, `conflicts' with 
the Dirichlet boundary
condition, and so, {\it is not} a physical  operator. This `deficiency' can be 
removed by the $PT$-symmetric regularization of  conformal mechanics systems $H_n$ via
a purely imaginary shift of the argument, $x\rightarrow x+i\alpha$,
$\R\ni  \alpha\neq 0$, accompanied by extension of $x$ from the half-line to the whole
real line. 
As it was shown in \cite{MatPly},
the resulting Hamiltonian 
\be\label{Halphan}
H_n^\alpha=-\frac{d^2}{dx^2}+\frac{n(n+1)}{(x+i\alpha)^2}\,,\qquad
x\in\R\,,
\ee
describes  a   perfectly invisible zero-gap $PT$-symmetric system, 
which is characterized by a \emph{purely real} spectrum in conformity
with general properties of such class of the systems \cite{BenRev,Most}. The peculiarity of the 
system (\ref{Halphan}), 
however, is that  
its transmission amplitude  is equal to one for all values of energy $E>0$,
and it has one bound state of zero energy given by a square-integrable
on $\R$ wave function $\psi_0^{(n)}(x)={1}/{(x+i\alpha)^n}$
generated from the eigenstate $\psi_0^{(0)}=1$ of zero energy of the free 
particle system: $\psi_0^{(n)}=\mathcal{A}^\alpha_n\psi_0^{(0)}$,
where $\mathcal{A}^\alpha_n=A_n^\alpha A_{n-1}^\alpha \ldots A_1^\alpha$,
$A^\alpha_{l}=\frac{d}{dx}-\frac{l}{x+i\alpha}$.
The Lax-Novikov integral
\be\label{NaxNovn}
{P}_n^\alpha=\mathcal{A}_n^\alpha P_0\mathcal{A}_n^{\alpha\#}\,,
\ee
where 
 $\mathcal{A}_n^{\alpha\#}=A_1^{\alpha\#} \ldots A_n^{\alpha\#}$,
$A_{l}^{\alpha\#}=-\frac{d}{dx}-\frac{l}{x+i\alpha}$,
detects the ground state $\psi_0^{(n)}(x)$ by annihilating it, 
and distinguishes the deformed
plane wave eigenstates $\psi^{\pm k}=\mathcal{A}_n^\alpha e^{\pm ikx}$ with $E=k^2>0$ in the continuous 
part of the spectrum, $P_n^\alpha \psi^{\pm k}=\pm k^{2n+1}\psi^{\pm k}$.
\vskip0.1cm

We pass over now to investigation of the effect of the presence  
of the  Lax-Novikov integrals on symmetries of the $PT$-regularized conformal 
mechanics systems $H^\alpha_n$  and of their $\mathcal{N}=2$
super-extended versions. To this aim we 
first consider the case of the simplest system $H_1^\alpha$,
and  then we study  its supersymmetric extension.

\section{Symmetries of the $H^\alpha_1$ system}\label{SecSymH1}

The generator of Galilean transformations 
$G_0=x-2t{P}_0=x+2it\frac{d}{dx}$
of the free quantum particle  $H_0=-\frac{d^2}{dx^2}$,
$x\in \R$,   
depends explicitly on the time parameter $t$, and
satisfies Heisenberg equation of motion of the form
$i\frac{d }{d t} G_0=i\frac{\partial G_0}{\partial t}-[H_0,G_0]=0$.
In correspondence with this property, $G_0$ is identified
as a dynamical integral of motion.
Two other dynamical integrals of motion of $H_0$ are 
generators of dilatations, 
${D}_0=\frac{1}{4}\{G_0,{P}_0\}=
-\frac{i}{2}\left(x\frac{d}{dx}+\frac{1}{2}\right)
-tH_0$,
and special conformal
transformations,  
$K_0=(G_0)^2=x^2-8t{D}_0-4t^2H_0$.
The integrals $G_0$, $D_0$ and $K_0$
are not translationally-invariant,  
$G_0(x+\tau)=G_0(x)+\tau$,
$\mathcal{D}_0(x+\tau)= \mathcal{D}_0(x)+\frac{1}{2}\tau {P}_0$, 
 $K_0(x+\tau)=K_0(x)+\frac{1}{2}\tau G_0(x)+\frac{1}{4}\tau^2$. 
 The set of integrals $H_0$,  ${P}_0$,
$G_0$, ${D}_0$ and $K_0$ 
and the unit operator  $\mathbb{I}$ 
generate the Schr\"odinger algebra  
\begin{eqnarray}\label{sl(2,R)}
&[{D}_0,H_0]=iH_0\,,\qquad
[{D}_0,K_0]=-iK_0\,,\qquad
 [K_0,H_0]=8i{D}_0\,,&\\
\label{D_0F}
&[{D}_0,{P}_0]=\frac{i}{2}{P}_0\,,\qquad 
[{D}_0,G_0]=-\frac{i}{2}G_0\,,&\\
\label{H_0F}
&[H_0,G_0]=-2i{P}_0\,,
\qquad
[H_0,{P}_0]=0\,,&\\
\label{K_0F}
&[K_0,{P}_0]=2iG_0\,,\qquad
[K_0,G_0]=0\,,&\\
\label{G_0P_0}
&[G_0,{P}_0]=i\,\mathbb{I} \,.&
\end{eqnarray}
This Lie algebra describes  symmetry of the free particle.
It  is the semi-direct sum of the conformal algebra $sl(2,\R)$ generated 
by $H_0$, $D_0$ and $K_0$, and of the one-dimensional
Heisenberg algebra generated by $P_0$, $G_0$ and $\mathbb{I}$.
The identity operator  $\mathbb{I}$ (in the chosen units $\hbar=1$, $m=1/2$) is the 
mass central element of the algebra.
According to Eqs. (\ref{D_0F}), (\ref{H_0F})  and 
(\ref{K_0F}),  the generators of translations, ${P}_0$, and 
Galileo transformations, $G_0$, form a doublet under the adjoint action 
of the  generators $H_0$, $K_0$ and ${D}_0$
of the conformal $sl(2,\R)$ symmetry.
The  algebra (\ref{sl(2,R)})--(\ref{G_0P_0}) 
is characterized  by the automorphism corresponding to a 
 spatial reflection,  
$ \rho_1: 
 {P}_0\rightarrow -{P}_0$, 
$G_0\rightarrow -G_0$, 
$H_0\rightarrow H_0$, 
$K_0\rightarrow K_0$, 
$D_0\rightarrow D_0$, 
$ \mathbb{I}\rightarrow \mathbb{I}$. 
It also has another automoprphism
\be\label{sigma_2H_0}
\rho_2:\,\,H_0\rightarrow K_0\,,\quad K_0\rightarrow H_0\,,\quad
D_0\rightarrow -D_0\,,\quad {P}_0\rightarrow -G_0\,,\quad
G_0 \rightarrow {P}_0\,,\quad \mathbb{I}\rightarrow \mathbb{I}\,,
\ee
which at $t=0$ corresponds to a unitary (canonical) transformation
$x\rightarrow -i\frac{d}{dx}$, $-i\frac{d}{dx} \rightarrow -x$.

\vskip0.1cm
Let us look now for extension 
and generalization of the Schr\"odinger  Lie algebra
(\ref{sl(2,R)})--(\ref{G_0P_0})
for the case of the $PT$-regularized 
conformal mechanics model $H^\alpha_1$.
For this we first identify formally
the integrals for the 
 conformal mechanics model $H_1$
by Darboux-dressing the integrals of  $H_0$
and considering their commutation relations.
Only generators of the conformal $sl(2,\R)$ symmetry
identified in such a way will be true integrals of motion
of $H_1$, while other formal integrals will be 
non-physical\,:  acting on physical eigentstates
of $H_1$  satisfying the Dirichlet boundary condition
$\psi(0)=0$, they produce non-physical states
satisfying the Neumann boundary condition $(\frac{d}{dx}\psi)(0)=0$.
The subsequent shift $x\rightarrow x+i\alpha$
(accompanied by extension $x>0\rightarrow x\in\R$
and omission of boundary condition for  wave functions at $x=0$)
 transforms then all 
the true and formal integrals of $H_1$
into the true integrals of motion of the $PT$-regularized 
conformal mechanics system $H^\alpha_1$.
 The key point  also is that here the substitution 
 $x\rightarrow x+i\alpha$,
 $\frac{d}{dx}\rightarrow \frac{d}{dx}$
 changes the  operators
 but does not touch the form of all the corresponding 
 nonlinear algebraic relations they satisfy.
 \vskip0.1cm

Similarly to a formal (non-physical) integral $P_1$
obtained via the  Darboux-dressing  of the
free particle momentum, we identify 
the analog of $G_0$  to be
a  differential operator
\be\label{G1(x)}
G_1(x)\equiv 
A_1(x)G_0(x)A_1^\#(x)=x\frac{d}{dx}A_1^\#(x)-2t{P}_1(x)\,.
\ee
Here the argument in 
integrals ${P}_1(x)$ and $G_1(x)$ 
is shown to stress 
that they are obtained 
from the corresponding integrals of the free particle $H_0$
via  Darboux-dressing by the intertwining 
operators $A_1(x)=\frac{d}{dx}-\frac{1}{x}$ and 
$A_1^\#(x)=-\frac{d}{dx}-\frac{1}{x}$.
The analog of ${D}_0(x)$ for the system $H_1(x)=A_1(x)A_1^\#(x)$ is
\begin{eqnarray}
&{D}_1(x)=-\frac{i}{2}\left(x\frac{d}{dx}+\frac{1}{2}\right)
-tH_1(x)\,.&
\end{eqnarray}
It can be extracted from the Darboux-dressed form of ${D}_0(x)$
by using  the  relations
$A_1(x){D}_0(x)A^\#_1(x)=
\left({D}_1(x)-\frac{i}{2}\right)H_1(x)=
H_1(x)\left({D}_1(x)+\frac{i}{2}\right)$,  
Analogously,  
from the relations 
$A_1(x)K_0(x)A^\#_1(x)=K_1(x)H_1(x)-
4i{D}_1(x)-1=
H_1(x)K_1(x)+4i
{D}_1(x)-1$ 
we extract a dynamical integral 
\be\label{C1(x)}
K_1(x)=x^2-8t{D}_1(x)-4t^2H_1(x)\,.
\ee
The operators ${D}_1$, $K_1$ and $H_1$ 
generate the $sl(2,\R)$ algebra
of the form (\ref{sl(2,R)})\,:
\be\label{sl(2,R)H_1}
[{D}_1,H_1]=iH_1\,,\qquad
[{D}_1,K_1]=-iK_1\,,\qquad
 [K_1,H_1]=8i{D}_1\,.
\ee 
The commutation relations
\[
[H_1,G_1]=-2i{P}_1\,,\qquad
[H_1, {P}_1]=0\,,
\]
are a direct analog of    (\ref{H_0F}),
while relations (\ref{D_0F}) and the first relation from (\ref{K_0F}) are replaced by
\begin{eqnarray}
\label{D_1F}
&[{D}_1,{P}_1]=\frac{3}{2}i{P}_1\,,\qquad
[{D}_1,G_1]=\frac{i}{2}G_1\,,&\\
&[K_1,{P}_1]=6iG_1\,.&
\end{eqnarray}
Instead of (\ref{G_0P_0}) 
we have a nonlinear commutation relation
\be\label{G_1P_1}
[G_1,{P}_1]=3i(H_1)^2\,.
\ee
The zero commutator $[K_0,G_0]=0$ is changed  for nonzero one,
\be\label{Lambda}
[K_1,G_1]=-4iV_1\,,
\ee
where 
\be
V_1(x)=ix^2A^\#_1(x)-4tG_1(x)-4t^2{P}_1(x)\,
\ee
is identified as a new explicitly depending on time 
formal integral of motion for the system $H_1$.
It  has the following commutation relations
with other integrals\,:
\begin{eqnarray}
&[V_1,H_1]=4iG_1\,,\qquad
[V_1,{D}_1]=\frac{i}{2}V_1\,,\qquad
[V_1,K_1]=2iR_1\,,&\\
\label{V_1quad}
&[V_1,{P}_1]=12iH_1{D}_1-6H_1\,,\qquad
[V_1,G_1]=12i ({D}_1)^2 +\frac{3}{4}i\, \mathbb{I} \,.&
\end{eqnarray}
 Here the operator
\be
R_1(x)=x^3-6tV_1(x)-12t^2G_1(x)-8t^3{P}_1
\ee
has to be identified as yet another new  formal  dynamical integral of motion
of $H_1$.
Its commutation relations with the rest of the integrals are
\begin{eqnarray}
&[R_1,H_1]=6iV_1\,,\qquad
[R_1,{D}_1]=\frac{3}{2}iR_1\,,
\quad
[R_1,K_1]=0\,,&\\
\label{R_1quad}
&[R_1,{P}_1]=36i\,{D}_1^2+\frac{21}{4}i\, \mathbb{I} \,,\quad
[R_1,G_1]=12i\,{D}_1K_1-6K_1\,,\quad
[R_1,V_1]=3i\,K_1^2\,.&
\end{eqnarray}
It may be noted here that the commutation relations of the form 
similar to (\ref{G_1P_1}) are satisfied by  ladder operators in rationally deformed 
harmonic oscillator systems as well as in rationally deformed
conformal mechanics model of de Alfaro-Fubini-Furlan with the included 
confining  harmonic potential term \cite{CarPly,CarInzPly}.
\vskip0.1cm

The explicitly depending on time 
\emph{formal}  integrals of $H_1$ have been 
 identified via the Darboux-dressing of the corresponding 
integrals of the free particle with additional step of subsequent `extraction'
in the case of the integrals ${D}_1$ and ${K}_1$.
The dynamical  integrals  also can be obtained via the  ``time-dressing" 
by the evolution operator $U_1(t)=\exp(iH_1t)$.
For this we note that the dynamical integral $X_1(t)=U_1^{-1}(t)xU_1(t)$
is given by an infinite series in $t$ and $\frac{d}{dx}$, and so, 
 is a \emph{nonlocal} in $x$  operator.
This is essentially different from  the free particle case 
where $U_0(t)=\exp(iH_0t)$ and $U_0^{-1}(t)xU_0(t)=G_0$ 
is the local operator $G_0$. 
However, 
the time dressing of the operators $x^2$,   
$\frac{1}{4}\{x,-i\frac{d}{dx}\}$, 
$x\frac{d}{dx}A_1^\#$,
$ix^2A^\#_1(x)$ and $x^3$
generates the local operators to be exactly the
dynamical integrals of motion ${K}_1(x)$, 
${D}_1(x)$, $G_1(x)$, $V_1(x)$ and $R_1(x)$,
respectively.
\vskip0.1cm

The integrals ${P}_1$ and $G_1$, being 
Darboux-dressed  free particle's 
generators of translations, ${P}_0$,
and Galileo transformations, $G_0$, 
are the third order  differential operators~\footnote{The 
not-depending explicitly on $t$ term in  (\ref{G1(x)})
is the second order differential operator.}. 
According to (\ref{D_0F}) and  (\ref{D_1F}),
the  scaling dimensions  $-1/2$ and $+1/2$ of 
${P}_0$ and $G_0$
 given by relation $i[{D}_{0},F]=s_F F$,  are changed here 
 for the scaling dimensions $-3/2$ and $-1/2$ of ${P}_1$ and $G_1$
 given by analogous relation with ${D}_{0}$ changed for
 ${D}_{1}$.
 Coherently with this, 
 the central charge in commutation relation (\ref{G_0P_0})
 is changed in  (\ref{G_1P_1}) for the operator  $(H_1)^2$  
 having  the scaling dimension  $-2$. 
 In addition, 
 two  new formal dynamical integrals  $V_1$ and $R_1$ 
 of the scaling dimensions $+1/2$ and $+3/2$ are generated
 via the commutation of $G_1$ with
  generator of the special conformal transformations 
  $K_1$ having the scaling dimension $+1$.
As a result,  instead of the Lie algebraic structure of 
the Schr\"odinger symmetry
of the free particle we obtain
the nonlinear algebra in which the commutators 
(\ref{G_1P_1}), (\ref{V_1quad}) and (\ref{R_1quad})
are quadratic in generators of the $sl(2,\R)$ Lie subalgebra
(\ref{sl(2,R)H_1}).
Instead of the doublet of integrals $(G_0, {P}_0)$
under the  adjoint action of the operators $H_0$, $K_0$
and ${D}_0$
in the case of the free particle, 
here we have the quartet $(R_1,V_1,G_1, {P}_1)$
under the adjoint action of 
the $sl(2,\R)$ generators  $H_1$, $K_1$
and ${D}_1$.

The described nonlinear 
algebra is characterized by the automorphism corresponding 
to a spatial reflection $\rho_1$.   
It also has the  automorphism $\rho_2$, 
\be\label{sigma_2H_1}
\rho_2:\,\,H_1\rightarrow K_1\rightarrow H_1\,,\,\,
D_1\rightarrow -D_1\,,\,\, {P}_1\rightarrow -R_1\,,\,\,
R_1 \rightarrow  {P}_1\,,\,\,
V_1\rightarrow -G_1\,, \,\,
G_1\rightarrow V_1\,.
\ee

Till the moment we  discussed the integrals of the system $H_1$.
Except the generators $H_1$, $D_1$ and $K_1$ of the $sl(2,\R)$ symmetry,
the rest of them are formal integrals being non-physical operators. 
By shifting  the  argument $x\rightarrow x+i\alpha$
and extending $x>0$ for $x\in\R$,
 we 
obtain the corresponding set of the true  integrals 
$H^\alpha_1$, $D^\alpha_1$, $K^\alpha_1$,
$P^\alpha_1$, $G^\alpha_1$, $V^\alpha_1$ and $R^\alpha_1$ 
of the $PT$-regularized conformal mechanics system $H^\alpha_1$.
They satisfy the nonlinearly extended Schr\"odinger algebra
of the same described form. 
\section{Symmetries of the $\mathcal{N}=2$ 
super-extended ${H}^\alpha_1$ system}\label{SecSusyExact}

Consider now the extended system described by the diagonal matrix
Hamiltonian operator 
\be\label{Hextend1x}
\mathcal{H}^\alpha=\text{diag}\, (H_1^\alpha,\, H_0)
\ee
composed from the $PT$-regularized 
conformal mechanics Hamiltonian  $H_1^\alpha$ and the free particle 
Hamiltonian $H_0$.
To simplify notations, below we omit the upper index 
$\alpha$ in the  Hamiltonian and matrix integrals
of the system (\ref{Hextend1x}).
The system (\ref{Hextend1x}) is described by the
superpotential $\mathcal{W}_e=-1/(x+i\alpha)$,
$\mathcal{H}=-\frac{d^2}{dx^2}+ \mathcal{W}_e^2-\mathcal{W}'_e\sigma_3$, 
$\mathcal{W}'_e=\frac{d}{dx}\mathcal{W}_e$, 
and is 
 characterized by the supercharges 
 \be\label{Qa}
\mathcal{Q}_1=
\left(
\begin{array}{cc}
 0 & A_1^\alpha   \\
A^{\alpha\#}_1  &  0
\end{array}
\right),\qquad
\mathcal{Q}_2=i\sigma_3\mathcal{Q}_1\,,
\ee
which are the matrix first order differential operators,
where $A_1^{\alpha}=\frac{d}{dx}+\mathcal{W}_e(x)$,
$A_1^{\alpha\#}=-\frac{d}{dx}+\mathcal{W}_e(x)$.
The peculiarity of the system (\ref{Hextend1x})
is that besides the generators of the $\mathcal{N}=2$ 
supersymmetry $\mathcal{Q}_a$, $a=1,2$,
it also possesses two more  supercharges
\be\label{Sa}
\mathcal{S}_1=
\left(
\begin{array}{cc}
 0 & -iA_1^\alpha P_0  \\
iP_0 A^{\alpha\#}_1  &  0
\end{array}
\right),\qquad
\mathcal{S}_2=i\sigma_3\mathcal{S}_1\,.
\ee
The appearance
of additional  supercharges (\ref{Sa}) to be matrix differential operators of the second order, 
as it clear from their structure, is explained by existence 
of the momentum integral $P_0$
in the free particle system. As a consequence,
the Hamiltonian operator $H_1^\alpha$  can be intertwined 
with the free particle Hamiltonian $H_0$ not only by the first order
operators $A_1^\alpha$ and $A^{\alpha\#}_1$ but also
by the second order operators from which 
supercharges (\ref{Sa}) are composed.
These four supercharges  satisfy the relations
\begin{eqnarray}
\label{susyHQS}
&[\mathcal{H},\mathcal{Q}_a]=0\,,\qquad
[\mathcal{H},\mathcal{S}_a]=0\,,&\\
\label{susyQQSS}
&\{\mathcal{Q}_a,\mathcal{Q}_b\}=2\delta_{ab}\mathcal{H}\,,\qquad
\{\mathcal{S}_a,\mathcal{S}_b\}=2\delta_{ab}\mathcal{H}^2\,,&\\
\label{susyQS}
&\{\mathcal{Q}_a,\mathcal{S}_b\}=2\epsilon_{ab}\mathcal{L}\,,&\\
\label{susyLHQS}
&[\mathcal{L},\mathcal{H}]=[\mathcal{L},\mathcal{Q}_a]=[\mathcal{L},\mathcal{S}_a]=0\,,&
\end{eqnarray}
where
\be\label{L1def}
\mathcal{L}=\text{diag}\, ( {P}^\alpha_1,\, H_0{P}_0)
\ee
is the bosonic integral of motion composed from
the Lax-Novikov integral  $ {P}^\alpha_1$
of the subsystem $H^\alpha_1$ and
the momentum operator $P_0$  of the free particle 
subsystem $H_0$.  The integrals $\mathcal{H}$, $\mathcal{Q}_a$, 
$\mathcal{S}_a$ and $\mathcal{L}$
generate the exotic nonlinear $\mathcal{N}=4$ superalgebra
of the system (\ref{Hextend1x}), in which 
the operator $\mathcal{L}$
plays a role of the bosonic central 
charge.
System (\ref{Hextend1x}) has 
the unique bound state of zero energy given by a square-integrable
on $\R$ wave function of the form $\Psi^t=({(x+i\alpha)^{-1}},0)^t$,
which is annihilated by all the superacharges 
$\mathcal{Q}_a$ and $\mathcal{S}_a$
as well as by the Lax-Novikov integral
$\mathcal{L}$.

We note here that the extension of 
the $\mathcal{N}=2$ supersymmetry up to the exotic 
nonlinear $\mathcal{N}=4$ supersymmetry 
in general case of the quantum reflectionless  and finite-gap systems
also is based on existence of the two pairs of  intertwiners 
to be differential operators of the even and odd orders.
In finite-gap systems, however, no analog
of the free particle system  with its proper 
integral of motion $P_0$ does appear.  
In another way the extension can be related to a presentation 
of the Lax-Novikov integral in such systems
in the form of the product of two non-singular operators of the 
even and odd differential orders. 
 For the detailed discussion
of these aspects see refs. \cite{CorJakPly,CorJakNP,AraMatPly,Defects1,Defects2,{ArPlcrys}}, where 
also an important  phenomenon of reduction of higher  order intertwining operators
to intertwining operators of a lower order  
is discussed. We consider some example of the  reduction 
below in Section \ref{spontbreake}.
Because of such a reduction mechanism,
each finite-gap or reflectionless system, including the $PT$-regularized 
conformal mechanics systems (\ref{Halphan}),
is  characterized by a corresponding unique irreducible Lax-Novikov integral\,:
reducible intertwining operators generate the same Lax-Novikov integral 
up to multiplication by a polynomial in the system's Hamiltonian.

The Hamiltonian  (\ref{Hextend1x}) 
together with the matrix dynamical bosonic integrals 
$\mathcal{D}=\text{diag}\,({D}_1^\alpha, {D}_0^\alpha)$
and $\mathcal{K}=\text{diag}\,({K}_1^\alpha, {K}_0^\alpha)$
generates the conformal algebra $sl(2,\R)$,
\be\label{Sch3+ext}
[\mathcal{D},\mathcal{H}]=i\mathcal{H}\,,\qquad
[\mathcal{D},\mathcal{K}]=-i\mathcal{K}\,,\qquad
 [\mathcal{K},\mathcal{H}]=8i\mathcal{D}\,,
\ee
where we use the notation $D^\alpha_0=D_0(x+i\alpha)$,
$K^\alpha_0=K_0(x+i\alpha)$.
Extending  this set of the bosonic integrals 
with the supercharge operators $\mathcal{Q}_a$ and 
$\mathcal{S}_a$, and
taking all the (anti)commutators of these operators
and the new integrals generated in this procedure,
we obtain a  nonlinear 
superalgebra which corresponds to some
nonlinear extension of the super-Schr\"odinger algebra.
It is generated by the set of the bosonic integrals 
$\mathcal{H}$, 
$\mathcal{D}$, $\mathcal{K}$,
$\mathcal{L}$, $\mathcal{G}$, 
$\mathcal{V}$, $\mathcal{R}$, $\mathcal{P}_-$,
$\mathcal{G}_-$, $\Sigma=\sigma_3$,
$\mathcal{I}=\text{diag}\,(1,1)$,
and by  the    fermionic   integrals 
$\mathcal{Q}_a$, $\mathcal{S}_a$,  and 
 $\lambda_a$, $\mu_a$ and $\kappa_a$, $a=1,2$,
where
\begin{eqnarray}\label{G(x)}
&\mathcal{G}=\text{diag}\,\left(G_1^\alpha,\,  \frac{1}{2}\{G_0^\alpha,H_0\}\right)\,,&
\end{eqnarray}
\be\label{Lambda}
\mathcal{V}=
i(x+i\alpha)^2A_1^{\alpha\#}\mathcal{I}-4t\mathcal{G}-4t^2\mathcal{L}\,,
\ee
 \be\label{mathcalR}
\mathcal{R}=
(x+i\alpha)^3\mathcal{I} -6t\mathcal{V} -12t^2\mathcal{G} -8t^3\mathcal{L}\,,
\ee
\begin{eqnarray}\label{mathP-}
&\mathcal{P_-}=\frac{1}{2}(1-\sigma_3){P}_0\,,\qquad
\mathcal{G_-}=\frac{1}{2}(1-\sigma_3){G}_0^\alpha\,,&
\end{eqnarray}
 \be\label{lambda}
\lambda_1=
\left(
\begin{array}{cc}
 0& i(x+i\alpha)  \\
-i(x+i\alpha)  &  0
\end{array}
\right)
 -2t\mathcal{Q}_1\,,\qquad \lambda_2=i\sigma_3\lambda_1\,,
\ee
 \be\label{mu1}
\mu_1=
\left(
\begin{array}{cc}
 0& (x+i\alpha) {P}_0 \\
{P}_0 (x+i\alpha)  &  0
\end{array}
\right) -2t\mathcal{S}_1\,,\qquad \mu_2=i\sigma_3\mu_1\,,
\ee
\be\label{kappa}
\kappa_1=
\left(
\begin{array}{cc}
 0& (x+i\alpha)^2  \\
(x+i\alpha)^2  &  0
\end{array}
\right)
 -4t\mu_1-4t^2\mathcal{S}_1\,,\qquad \kappa_2=i\sigma_3\kappa_1\,.
\ee
 The dynamical integrals represent 
  the 
 corresponding  time-independent operators   
 ``dressed"  
  by the matrix evolution operator 
 $\mathcal{U}(t)=\exp(i\mathcal{H}t)$\,:
 $F(t)=\mathcal{U}^{-1}(t) F(0) \mathcal{U}(t)$.
 Particularly,   the dynamical integral $\mathcal{G}$ 
 corresponds to a dressed form of the diagonal matrix operator
 $g=\text{diag}\,\left((x+i\alpha)\frac{d}{dx}A^{\alpha\#}_1,\, 
 (x+i\alpha)A^{\alpha\#}_1\frac{d}{dx}\right)$.
 
 All the integrals  are 
 eigenstates of the operator $\mathcal{D}$ in the sense of its 
 adjoint action\,:
 \begin{eqnarray}\label{Dcom1}
& 
 [\mathcal{D},\mathcal{I}]=[\mathcal{D},\mathcal{D}]=[\mathcal{D},\Sigma]=
 [\mathcal{D},\mu_a]=0\,,&\\
 & [\mathcal{D},\mathcal{H}]=i\mathcal{H}\,,\quad
  [\mathcal{D},\mathcal{K}]=-i\mathcal{K}\,,&\\
&  [\mathcal{D},\mathcal{L}]=\frac{3}{2}i\mathcal{L}\,,\qquad
     [\mathcal{D},\mathcal{R}]=-\frac{3}{2}i\mathcal{R}\,,&\\
& [\mathcal{D},\mathcal{G}]=\frac{i}{2}\mathcal{G}\,,\quad
  [\mathcal{D},\mathcal{V}]=-\frac{i}{2}\mathcal{V}\,,\quad
    [\mathcal{D},\mathcal{P}_-]=\frac{i}{2}\mathcal{P}_-\,,\quad
       [\mathcal{D},\mathcal{G}_-]=-\frac{i}{2}\mathcal{G}_-\,,&\\
 &   [\mathcal{D},\mathcal{Q}_a]=\frac{i}{2}\mathcal{Q}_a\,,\quad
 [\mathcal{D},\mathcal{S}_a]=i\mathcal{S}_a\,,\quad
          [\mathcal{D},\lambda_a]=-\frac{i}{2}\lambda_a\,\quad
           [\mathcal{D},\kappa_a]=-i\kappa_a\,.&\label{SQDLMK}
\end{eqnarray}
The complete set of other  (anti)-commutation relations is\,:
\begin{eqnarray}
& [\mathcal{H},\mathcal{K}]=-8i\mathcal{D}\,,\qquad
 [\mathcal{H},\mathcal{D}]=-i\mathcal{H}\,,&\\
 & [\mathcal{H},\mathcal{R}]=-6i\mathcal{V}\,,\quad
   [\mathcal{H},\mathcal{V}]=-4i\mathcal{G}\,,\quad
 [\mathcal{H},\mathcal{G}]=-2i\mathcal{L}\,,\quad
  [\mathcal{H},\mathcal{L}]=0\,,&\\
 & [\mathcal{H},\mathcal{G}_-]=-2i\mathcal{P}_-\,,\qquad
   [\mathcal{H},\mathcal{P}_-]=0\,,&\\
& [\mathcal{H},\mathcal{Q}_a]=0\,,\qquad
 [\mathcal{H},\mathcal{S}_a]=0\,,&\\
& [\mathcal{H},\kappa_a]=-4i\mu_a\,,\qquad
 [\mathcal{H},\mu_a]=-2i\mathcal{S}_a\,,\qquad
 [\mathcal{H},\lambda_a]=-2i\mathcal{Q}_a\,,&
 \end{eqnarray}
 \begin{eqnarray}\label{LKG}
 &   [\mathcal{K},\mathcal{L}]=6i\mathcal{G}\,,\quad
     [\mathcal{K},\mathcal{G}]=4i\mathcal{V}\,,\quad
      [\mathcal{K},\mathcal{V}]=2i\mathcal{R}\,,\quad
         [\mathcal{K},\mathcal{R}]=0\,,&\\
&    [\mathcal{K},\mathcal{P}_-]=2i\mathcal{G}_-\,,\qquad
      [\mathcal{K},\mathcal{G}_-]=0\,,&\\
 &  [\mathcal{K},\mathcal{Q}_a]=2i\lambda_a\,,
  \quad
      [\mathcal{K},\lambda_a]=0\,,\quad
    [\mathcal{K},\mathcal{S}_a]=4i\mu_a\,,\quad
        [\mathcal{K},\mu_a]=2i\kappa_a\,,\quad
          [\mathcal{K},\kappa_a]=0\,,&\label{KQaSalama}
  \end{eqnarray}
 \begin{eqnarray}
&   [\mathcal{L},\mathcal{G}]=-3i\mathcal{H}^2\,,\qquad
     [\mathcal{L},\mathcal{P}_-]=0\,,\qquad
       [\mathcal{L},\mathcal{G}_-]=-3i\Pi_-\mathcal{H}\,,&\\
  & [\mathcal{L},\mathcal{V}]=-3\left(1+\Upsilon^{+}\right) \mathcal{H}
   \,,\qquad
     [\mathcal{L},\mathcal{R}]=-i\left(
     36\mathcal{D}^2+\frac{9}{2}\Sigma +\frac{3}{4}\mathcal{I}\right),&\\
&   [\mathcal{L},\mathcal{Q}_a]=0\,,\qquad
   [\mathcal{L},\mathcal{S}_a]=0 \,,&\\
 &    [\mathcal{L},\lambda_a]=3i\epsilon_{ab}\mathcal{S}_b\,,\quad
       [\mathcal{L},\mu_a]=3i\mathcal{H}\epsilon_{ab}\mathcal{Q}_b\,,
       \quad
        [\mathcal{L},\kappa_a] =3\big(\Upsilon^+\epsilon_{ab}+i\delta_{ab}\big)\mathcal{Q}_b\,,&
       \end{eqnarray}
 \begin{eqnarray}
 & [\mathcal{G},\mathcal{V}]=3i\mathcal{H}\mathcal{K}-12\mathcal{D}+\frac{i}{2}
 (7 \Sigma-\mathcal{I})\,,\qquad
  [\mathcal{G},\mathcal{R}]=3(1+\Upsilon^-)\mathcal{K}\,,&\\
 & [\mathcal{G},\mathcal{P}_-]=i\Pi_-\mathcal{H}\,,\qquad
  [\mathcal{G},\mathcal{G}_-]=-4i\Pi_-\mathcal{D}\,,&\\
&    [\mathcal{G},\mathcal{Q}_a]=i\epsilon_{ab}\mathcal{S}_b\,,\qquad
    [\mathcal{G},\mathcal{S}_a]=-2i\mathcal{H}\epsilon_{ab}\mathcal{Q}_b\,,&\\
    & [\mathcal{G},\lambda_a]=-2i\epsilon_{ab}\mu_b\,,&\\
      & [\mathcal{G},\mu_a]=\frac{1}{2}\Upsilon^+\epsilon_{ab}\mathcal{Q}_b\,,
       \qquad
        [\mathcal{G},\kappa_a]=-\big(2\Upsilon^- \epsilon_{ab}+i\delta_{ab}\big)
        \lambda_b,&
    \end{eqnarray}
  \begin{eqnarray}
 & [\mathcal{V},\mathcal{R}]=-3i\mathcal{K}^2\,,\qquad
  [\mathcal{V},\mathcal{P}_-]=4i\Pi_-\mathcal{D}\,,\qquad
   [\mathcal{V},\mathcal{G}_-]=-i\Pi_-\mathcal{K}\,,&\\
 & [\mathcal{V},\mathcal{Q}_a]=2i\epsilon_{ab}\mu_b\,,\qquad
  [\mathcal{V},\mathcal{S}_a]=-\big(2\Upsilon^+\epsilon_{ab}+i\delta_{ab}\big)\mathcal{Q}_b\,,&\\
    &
  [\mathcal{V},\lambda_a]=-i\epsilon_{ab}\kappa_b\,,\quad
  [\mathcal{V},\mu_a]=\frac{1}{2}\Upsilon^- \epsilon_{ab}\lambda_b\,,\quad
  [\mathcal{V},\kappa_a]=4i\mathcal{K}\epsilon_{ab}\lambda_b\,,
  &
    \end{eqnarray}
\begin{eqnarray}
&    [\mathcal{R},  \mathcal{P}_-]=3i\Pi_-\mathcal{K}\,,\qquad
      [\mathcal{R},  \mathcal{G}_-]=0\,,&\\
  &    [\mathcal{R},\mathcal{Q}_a]=-3i\epsilon_{ab}\kappa_b\,,\qquad
         [\mathcal{R},\mathcal{S}_a]=3\big(\Upsilon^- \epsilon_{ab}+i\delta_{ab}\big)\lambda_b\,,&\\
 & [\mathcal{R},  \lambda_a]=0\,,\qquad
     [\mathcal{R},\mu_a]=-3i\mathcal{K}\epsilon_{ab}\lambda_b\,,\qquad
        [\mathcal{R},\kappa_a]=0\,,&
\end{eqnarray}
 \begin{eqnarray}
 &    [\mathcal{P}_-,   \mathcal{G}_-]=-\frac{i}{2}(\mathcal{I}-\Sigma)\,,\qquad
     [\mathcal{P}_-,\mathcal{Q}_a]=-i\mathcal{S}_a\,,\qquad
        [\mathcal{P}_-,\mathcal{S}_a]=i\mathcal{H}\mathcal{Q}_a\,,&\\
   & [\mathcal{P}_-,\lambda_a]=-i\mu_a\,,&\\
     &[\mathcal{P}_-,\mu_a]=\frac{1}{2} \Upsilon^+  \mathcal{Q}_a\,,\qquad
      [\mathcal{P}_-,\kappa_a]=\left(-\frac{1}{2}\Upsilon^- \delta_{ab}+
       i\epsilon_{ab}\right)\lambda_b\,,&
  \end{eqnarray}
     \begin{eqnarray}
    & [\mathcal{G}_-,\mathcal{Q}_a]=-i\mu_a\,,\qquad
        [\mathcal{G}_-,\mathcal{S}_a]= 
        -\mathcal{Q}_a+i\mathcal{H}\lambda_a\,,&
        \\
  &  [\mathcal{G}_-,\lambda_a]=-i\kappa_a\,,\qquad
     [\mathcal{G}_-,\mu_a]=-\frac{1}{2}\Upsilon^- \lambda_a\,,\qquad
      [\mathcal{G}_-,\kappa_a]=i\mathcal{K}\lambda_a\,,&
  \end{eqnarray}
      \begin{eqnarray}
&  [\Sigma,\mathcal{B}]=0\,,\qquad \mathcal{B}=\Gamma, \mathcal{H}, \mathcal{L},
  \mathcal{D},
  \mathcal{K},\mathcal{G},\mathcal{V},\mathcal{R},\mathcal{P}_-,
  \mathcal{G}_-\,,&\\
   & [\Sigma,\mathcal{F}_a]=2i\epsilon_{ab}\mathcal{F}_b\,,\qquad
   \mathcal{F}_a=\mathcal{Q}_a,\mathcal{S}_a,\lambda_a,\mu_a,\kappa_a\,, &
   \label{SigmaF}
  \end{eqnarray}
    \begin{eqnarray}
& \{\mathcal{Q}_a,\mathcal{Q}_b\}=2\delta_{ab}\mathcal{H}\,,\qquad
 \{\mathcal{Q}_a,\mathcal{S}_b\}=2\epsilon_{ab}\mathcal{L}\,,&\\
  & \{\mathcal{Q}_a,\lambda_b\}=4\delta_{ab}\mathcal{D}
   +2\epsilon_{ab}\big(\mathcal{I}-\frac{1}{2}\Sigma \big)\,, &\\
   & \{\mathcal{Q}_a,\mu_b\}=-2\delta_{ab}\mathcal{P}_-+2\epsilon_{ab}
    \mathcal{G}\,,\qquad
     \{\mathcal{Q}_a,\kappa_b\}=-4\delta_{ab}\mathcal{G}_-+2\epsilon_{ab}\mathcal{V}\,,&
     \end{eqnarray}
     \begin{eqnarray}
& \{\mathcal{S}_a,\mathcal{S}_b\}=2\delta_{ab}\mathcal{H}^2\,,&\\
 &\{\mathcal{S}_a,\lambda_b\}=4\delta_{ab}\mathcal{P}_--2\epsilon_{ab}\mathcal{G}\,,\qquad
  \{\mathcal{S}_a,\mu_b\}=\big(-i(1+\Upsilon^+)\delta_{ab} +(1-2\Sigma)\epsilon_{ab}\big)\mathcal{H}\,,&\\
   &
  \{\mathcal{S}_a,\kappa_b\}=\left(\frac{1}{2}\mathcal{H}\mathcal{K}-\mathcal{I}\right)\delta_{ab}+\big(2i
 \delta_{ab} +4(1-2\Sigma)\epsilon_{ab}\big) \mathcal{D}\,,&
   \end{eqnarray}
     \begin{eqnarray}
& \{\lambda_a,\lambda_b\}=2\delta_{ab}\mathcal{K}\,,\qquad
  \{\lambda_a,\mu_b\}=2\epsilon_{ab}\mathcal{V}
  +2\delta_{ab}\mathcal{G}_-\,,\qquad
   \{\lambda_a,\kappa_b\}=2\epsilon_{ab}\mathcal{R}\,,&\\
    &
    \{\mu_a,\mu_b\}=\frac{1}{2}\Upsilon^+ \Upsilon^- \delta_{ab}
    \,,\qquad
     \{\mu_a,\kappa_b\}=\big(i(1+\Upsilon^-) \delta_{ab}
     +(1-2\Sigma)\epsilon_{ab}\big)\mathcal{K}\,,
    &\\
  & \{\kappa_a,\kappa_b\}=2\delta_{ab}\mathcal{K}^2\,.&\label{kakaK}
      \end{eqnarray}
   We use here a notation $\Pi_-=\frac{1}{2}(1-\Sigma)$ and 
    $\Upsilon^\pm=1\pm4i\mathcal{D}$ for operator-valued coefficients appearing in 
    some (anti)-commutation relations.
   \vskip0.1cm
   
   The superalgebra (\ref{Dcom1})--(\ref{kakaK}) represents a  nonlinear generalization  
   of the super-Schr\"odinger symmetry
   with ten even and ten odd generators plus a bosonic central charge
   $\mathcal{I}$.
 It is characterized  by the 
 quadratic Casimir operator  
 $C_0=\{\mathcal{K},\mathcal{H}\}-8\mathcal{D}^2+
 2\left(\mathcal{I}-\frac{1}{2}\Sigma\right)^2-3\mathcal{I}^2$,
which takes zero value $C_0=0$
for the system (\ref{Hextend1x}).
 The set of operators $\mathcal{H}$,
$\mathcal{K}$, $\mathcal{D}$, $(\mathcal{I}-\frac{1}{2}\Sigma)$,
$\mathcal{Q}_a$ and $\lambda_a$ generates the Lie
sub-superalgebra $osp(2\vert 2)$ of superconformal
symmetry of the system. The extension of the 
indicated set of operators
by  the Lax integral $\mathcal{L}$ or by any other integral
not appearing in the list
produces all the
rest of the generators and expands the 
superalgebra $osp(2\vert 2)$ up to  the described 
nonlinear (quadratic) superalgebra. 
As an example, one such a chain of the (anti)commutation relations
can be presented schematically as
\be\nonumber
\mathcal{L}\overset{\mathcal{K}\,}{\longrightarrow} \mathcal{G}\overset{\mathcal{K}\,}{\longrightarrow}
\mathcal{V}\overset{\mathcal{K}\,}{\longrightarrow}\mathcal{R}\,,\qquad
\mathcal{L}\overset{{\lambda_a}\,}{\longrightarrow}
\mathcal{S}_a \overset{{\lambda_a}\,}{\longrightarrow} \mathcal{P}_- \overset{{\lambda_a}\,}{\longrightarrow}
\mu_a \overset{{\lambda_a}\,}{\longrightarrow}
\mathcal{G}_-  \overset{{\lambda_a}\,}{\longrightarrow}\kappa_a\,,
\qquad \mathcal{G}_-
\overset{{\mathcal{P}_-}\,}{\longrightarrow}\mathcal{I},\Sigma\,,
\ee
where $\mathcal{L}\overset{\mathcal{K}\,}{\longrightarrow} \mathcal{G}$
corresponds to the first commutation relation in  (\ref{LKG}), etc.

The peculiarity of  the obtained nonlinear superalegbra 
is that 
the (anti)-commutators of the generators of the 
$osp(2\vert 2)$ 
sub-superalgebra with any other generator  is linear in 
generators;
at the same time, in the (anti)-commutators 
of the rest of the  integrals, 
the generators $\mathcal{H}$,
$\mathcal{K}$, $\mathcal{D}$ and $\Sigma$ of the 
$sl(2,\R)\oplus u(1)$ sub-algebra
appear as operator-valued coefficients.

The  superalgebra  has the automorphism corresponding to
a spatial reflection $\rho_1$, under which the integrals  
  $\mathcal{L}$, $\mathcal{G}$,   $\mathcal{V}$,  $\mathcal{R}$, 
   $\mathcal{P}_-$,  $\mathcal{G}_-$, $\mathcal{Q}_a$, and $\lambda_a$
   are odd (change the sign), while the rest 
of generators  is even.
Another set of   transformations
\begin{eqnarray}
&\rho_2:\,\,  \mathcal{H}\rightarrow  \mathcal{K}\,,\quad
  \mathcal{K}\rightarrow  \mathcal{H}\,,\quad
  \mathcal{D}\rightarrow  -\mathcal{D}\,,\quad
  \mathcal{R}\rightarrow  \mathcal{L}\,,\quad 
    \mathcal{L}\rightarrow  -\mathcal{R}\,,&\nonumber\\
 &\mathcal{V}\rightarrow  -\mathcal{G}\,,\quad
  \mathcal{G}\rightarrow  \mathcal{V}\,,\quad
  \mathcal{G}_-\rightarrow  \mathcal{P}_-\,,\quad
  \mathcal{P}_-\rightarrow  -\mathcal{G}_-\,,\quad 
   {\Sigma}\rightarrow  \Sigma\,,\quad 
    \mathcal{I}\rightarrow  \mathcal{I}\,,&\nonumber\\
 &\mathcal{Q}_a\rightarrow  -\lambda_a\,,\quad
  \lambda_a\rightarrow  \mathcal{Q}_a\,,\quad
  \mathcal{S}_a\rightarrow  \kappa_a\,,\quad
  \kappa_a\rightarrow  \mathcal{S}_a\,,\quad 
  \mu_a \rightarrow  -\mu_a\,,&\nonumber
\end{eqnarray}
corresponds to automorphism of the superalgebra
that unifies and generalizes relations (\ref{sigma_2H_0})
and (\ref{sigma_2H_1}) for the superextended system
(\ref{Hextend1x}).

\section{Symmetries of $H^\alpha_n$ and of its super-extended version}\label{Section5}

In this section we generalize the analysis 
of the previous two sections for the case of
the $PT$-regularized conformal mechanics system 
$H^\alpha_n$ and its super-extended version.

System  (\ref{Halphan}) 
has the dynamical integrals of motion 
\begin{eqnarray}\label{Dalphan}
&D^\alpha_n=-\frac{i}{2}\left((x+i\alpha)\frac{d}{dx}+\frac{1}{2}\right)-tH^\alpha_n\,,
\qquad
K_n^\alpha=(x+i\alpha)^2-8tD_n^\alpha-4t^2H_n^\alpha\,,&
\end{eqnarray}
which together with  Hamiltonian $H_n^\alpha$
generate the conformal $sl(2,\R)$ symmetry.
It
also has the Lax-Novikov  integral of motion $P_n^\alpha$, $[P^\alpha_n,H^\alpha_n]=0$,
being the Darboux-dressed momentum integral $P_0$
of the free particle.
Its scale dimension is $-(n+\frac12)$,
$i[D^\alpha_n,P^\alpha_n]=-(n+\frac{1}{2})P^\alpha_n$.
Repeated commutation of $P_n^\alpha$ with $K_n^\alpha$,
$[P^\alpha_n,K_n^\alpha]$, $[[P_n^\alpha,K_n^\alpha],K_n^\alpha]$, $\ldots$,
generates in addition to $D^\alpha_n$ and $K^\alpha_n$, the  
$2n$ dynamical integrals  of the half-integer scale dimensions
$-(n-\frac{1}{2}),\ldots, (n+\frac{1}{2})$.
These are the operators 
which together with the  integral $P^\alpha_n$ can be presented in the form 
\be\label{Xcalnl}
\mathcal{X}_{n,\ell}^{\alpha}=U_n^{-1}(t){X}_{n,\ell}^{\alpha}U_n(t)={X}_{n,\ell}^{\alpha}+\ldots\,, 
\qquad
\ell=0,1,\ldots, 2n+1\,,
\ee
where $U_n(t)=\exp(itH_n^\alpha)$, the ellipsis corresponds to a  polynomial 
of order $\ell$ in $t$,
\begin{eqnarray}
&{X}^\alpha_{n,\ell}=(x+i\alpha)^\ell \mathcal{A}^\alpha_{n,\ell}\,,&\\
&\mathcal{A}^\alpha_{n,\ell}=A^\diamondsuit_{\ell-n}A^\diamondsuit_{\ell+1-n}\ldots A^\diamondsuit_n\,,\quad \ell=0,1,\ldots,2n\,,
\qquad
 \mathcal{A}^\alpha_{n,2n+1}=1\,,&
\end{eqnarray}
and  we denote  $A^\diamondsuit_{-k}=A^{\alpha}_k$ for $k>0$,
 $A^\diamondsuit_{0}=\frac{d}{dx}$, and $A^\diamondsuit_{k}=A^{\alpha\#}_{k}$ for $k>0$.
 In particular cases of $\ell=0$ and $\ell=2n+1$
 the integrals  (\ref{Xcalnl}) are reduced to $\mathcal{X}_{n,0}=iP^\alpha_n$ 
 and $\mathcal{X}_{n,2n+1}=U_n^{-1}(t)(x+i\alpha)^{2n+1}U_n(t)$.
 The integrals (\ref{Xcalnl}) form a  $2(n+1)$-dimensional representation 
 with respect to the adjoint action of the $sl(2,\R)$ generators,
 \begin{eqnarray}
 &[D_n^\alpha,\mathcal{X}^\alpha_{n,\ell}]=i(n-\ell+\frac{1}{2})\mathcal{X}^\alpha_{n,\ell}\,,&\\
&  [H_n^\alpha,\mathcal{X}^\alpha_{n,\ell}]=2\ell\, \mathcal{X}^\alpha_{n,\ell-1}\,,\qquad
  [K_n^\alpha,\mathcal{X}^\alpha_{n,\ell}]=2(2n+1-\ell) \mathcal{X}^\alpha_{n,\ell+1}\,.&
  \end{eqnarray}
 The commutation relations  of $\mathcal{X}^\alpha_{n,\ell}$ between themselves
 are reduced to the form
\be
[\mathcal{X}^\alpha_{n,\ell},\mathcal{X}^\alpha_{n,\ell'}]=
\mathcal{P}_{\ell,\ell'}^{(2n)}(H_n^\alpha,K_n^\alpha,D_n^\alpha)\,,
\ee 
 where $\mathcal{P}_{\ell,\ell'}^{(2n)}$ are some polynomials of order $2n$ in the 
 $sl(2,\R)$ generators.

The supersymmetric  system given by the superpotential
$\mathcal{W}_e=-n/(x+i\alpha)$
is described by the matrix Hamiltonian
\be\label{Hextendnx}
\mathcal{H}_n=\text{diag}\,( H_n^\alpha,\,  H^\alpha_{n-1})\,.
\ee
System (\ref{Hextendnx}) is 
characterized  by nonlinear extension of the Schr\"odinger superalgebra,
which in this case is generated by $4n+6$ bosonic integrals and 
the same number of fermionic integrals, and by the bosonic central charge $\mathcal{I}$.
The set of bosonic generators  is given by $\mathcal{H}_n$,
$\mathcal{D}_n=\text{diag}\,(D^\alpha_n, D^\alpha_{n-1})$, 
$\mathcal{K}_n=\text{diag}\,(K^\alpha_n, K^\alpha_{n-1})$, 
$\Sigma$,
$\mathcal{X}^+_{n,\ell}=\mathcal{X}_{n,\ell}^{\alpha}\Pi_+$, $\ell=0,\ldots,2n+1$,
and $\mathcal{X}^-_{n,k}=\mathcal{X}_{n-1,k}^{\alpha}\Pi_-$,
$k=0,\ldots,2n-1$, where $\Pi_\pm=\frac{1}{2}(1\pm\sigma_3)$.
The set of fermionic generators is 
$\mathcal{Q}_{n,a}$, $\lambda_{n,a}$, 
$\mathcal{S}_{n,a}$, 
and $\mu_{n,k,a}$, $k=0,\ldots,2n-1$. 
The explicit form of these generators with index $a=1$ is
given by
\begin{eqnarray}\label{lambda}
&\mathcal{Q}_{n,1}=
\left(
\begin{array}{cc}
 0& A^-_n  \\
A^\#_n  &  0
\end{array}
\right), \qquad
\lambda_{n,1}=
\left(
\begin{array}{cc}
 0& i(x+i\alpha)  \\
-i(x+i\alpha)  &  0
\end{array}
\right)-2t\mathcal{Q}_{n,1}\,, &\\
\label{lambda}
&\mathcal{S}_{n,1}=
\left(
\begin{array}{cc}
 0& -iA^\alpha_n P^\alpha_{n-1} \\
 iP^\alpha_{n-1}A^{\alpha\#}_n &  0
\end{array}
\right), &\\
\label{kappan}
&\mu_{n,k,1}=\mathcal{U}^{-1}_n(t)
\left(
\begin{array}{cc}
 0& (x+i\alpha)X^\alpha_{n-1,k}\\
(-1)^{k+1}X^\alpha_{n-1,k}(x+i\alpha) &  0
\end{array}
\right)\mathcal{U}_n(t) \,.&
\end{eqnarray}
The generators with index $a=2$ are obtained by
multiplication with $i\sigma_3$, $\mathcal{Q}_{n,2}=i\sigma_3\mathcal{Q}_{n,1}$,
etc., and  here $\mathcal{U}_n(t)=\exp(i\mathcal{H}_nt)$ is the
evolution operator.
The operators  $\mu_{n,k,a}$ with  $k=2n-1$ 
and $k=n-1$ 
correspond, respectively,  to the fermionic dynamical integrals (\ref{kappa}) 
and (\ref{mu1}) in the case of  $n=1$.

The operators 
$\mathcal{H}_n$,
$\mathcal{K}_n$, $\mathcal{D}_n$, $(\mathcal{I}-\frac{1}{2}\Sigma)$,
$\mathcal{Q}_{n,a}$ and $\lambda_{n,a}$ are generators of  the Lie
sub-superalgebra $osp(2\vert 2)$ of superconformal
symmetry of the system (\ref{Hextendnx}).
Extension of this set by any other integral 
(different from the central charge $\mathcal{I}$)
results in expansion of the $osp(2\vert 2)$ up to
the whole nonlinearly extended super-Schr\"odinger  
algebra. 
As in the $n=1$ case, the (anti)-commutators
of all the integrals with the $osp(2\vert 2)$ generators 
are linear in the generators of the 
extended super-Schr\"odinger algebra.
Particularly, the operators $\mathcal{X}^+_{n,\ell}$
and $\mathcal{X}^-_{n,k}$
have the  scaling dimensions $-(n-\ell+\frac{1}{2}$) and  $-(n-k-\frac{1}{2})$, respectively,
given by
the adjoint action of $i\mathcal{D}_n$, 
while the pairs of fermionic operators $S_{n,a}$ and $\mu_{n,k,a}$, $a=1,2$,
have the scaling dimensions $-n$ and $(k+1-n)$. 
In the (anti)-commutators  of additional 
generators with generators  of the $osp(2\vert 2)$ 
superconformal symmetry,  the structure coefficients  are certain polynomials 
of order $2n-1$ in  generators $\mathcal{H}$,
$\mathcal{K}$, $\mathcal{D}$ and $\Sigma$ of the 
$sl(2,\R)\oplus u(1)$ sub-algebra.
The Hamiltonian $\mathcal{H}_n$, the supercharges 
$\mathcal{Q}_{n,a}$ and $\mathcal{S}_{n,a}$, 
and the bosonic Lax-Novikov matrix integral 
taken in the form $\mathcal{L}_n=\mathcal{X}^+_{n,0}
+\mathcal{H}_n\mathcal{X}^-_{n,0}=\text{diag}\,
(P^\alpha_n,H^\alpha_{n-1}P^\alpha_{n-1})$
generate the exotic nonlinear $\mathcal{N}=4$  
super-Poincar\'e algebra of the form (\ref{susyHQS}),
(\ref{susyQQSS}), (\ref{susyQS}), (\ref{susyLHQS})
with the  unique difference in the anti-commutation relation between 
the higher order supercharges $\mathcal{S}_{n,a}$\,:
$\{\mathcal{S}_{n,a},\mathcal{S}_{n,b}\}=2\delta_{ab}(\mathcal{H}_n)^{2n}$.
As in the particular case of $n=1$, 
the exotic nonlinear $\mathcal{N}=4$  
supersymmetry of the system (\ref{Hextendnx}) 
is  unbroken\,: its unique ground state of 
zero energy $\Psi^t=({(x+i\alpha)^{-n}},0)^t$
 is annihilated by all the supercharges
$\mathcal{Q}_{n,a}$ and $\mathcal{S}_{n,a}$ as well as by the Lax-Novikov integral
 \cite{MatPly}.
Here, the operator 
$\mathcal{G}_n= \mathcal{X}^+_{n,1}+
\frac{1}{2}\{\mathcal{H}_n,\mathcal{X}^-_{n,1}\}=
\text{diag}\,(\mathcal{X}^\alpha_{n,1},\frac{1}{2}\{\mathcal{H}_{n-1},\mathcal{X}^\alpha_{n-1,1}\})$
is the analog of the integral (\ref{G(x)}) of the case $n=1$.

\section{Spontaneoulsy broken phase of the exotic SUSY}\label{spontbreake}

We  study here the case of the $PT$-regularized $g=2$ superconformal 
mechanics model  in the phase of the partially broken 
 exotic nonlinear $\mathcal{N}=4$ super-Poincar\'e symmetry.
 The symmetry of the system we investigate  is described by the same number of 
 bosonic and fermionic  generators  as  in the system 
 from  Section \ref{SecSusyExact}, 
 but the structure  of  nonlinear super-algebra they generate
 is essentially  different from that of the model 
 (\ref{Hextend1x}).  Nevertheless, its  
 symmetry is shown may be  related to 
 the nonlinearly extended super-Schr\"odinger 
symmetry of the system  (\ref{Hextend1x}) 
in the limit of the transition to the phase of the unbroken 
exotic nonlinear 
$\mathcal{N}=4$ supersymmetry.

 The system we consider is  generated via the 
$\mathcal{N}=2$ supersymmetric quantum mechanics construction
based on the superpotential 
$\mathcal{W}_b(x)=1/\xi_1-1/\xi_2+i\delta^{-1}$, where 
$\xi_j=x+i\alpha_j$, $j=1,2$,  $\R\ni \alpha_j\neq 0$,
 $\delta=\alpha_1-\alpha_2\neq 0$. 
 In the limit when the parameter $\alpha_1$ 
 (or $\alpha_2$) is sent to infinity, the superpotential 
 transforms into the superpotential $\mathcal{W}_e(x)$
 (or $-\mathcal{W}_e(x)$) of the system 
  (\ref{Hextend1x}).
 We have 
$\mathcal{W}_b^2-\mathcal{W}_b'={2}/{\xi_1^2}-\delta^{-2}$,
$\mathcal{W}_b^2+\mathcal{W}_b'={2}/{\xi_2^2}-\delta^{-2}$,
that allows us to introduce the first order differential operators 
$A_b=\frac{d}{dx}+\mathcal{W}_b(x)$ and 
$A_b^\#=-\frac{d}{dx}+\mathcal{W}_b(x)$
satisfying the factorization,
$A_b^\#A_b=H^{\alpha_1}_1-\delta^{-2}$, 
$A_bA_b^\#=H^{\alpha_2}_1-\delta^{-2}$, 
and the intertwining,
$A_bH^{\alpha_1}_1=H^{\alpha_2}_1A_b$,
$A_b^\#H^{\alpha_2}_1=H^{\alpha_1}_1A_b^\#$,
 relations, 
where $H^{\alpha_j}_1=-\frac{d^2}{dx^2}+2/\xi^2_j $.
Define now the extended system
described by the matrix Hamiltonian 
\be\label{Hextendbr}
\mathcal{H}=\text{diag}\,( H^{\alpha_2}_1,\, H^{\alpha_1}_{1})\,.
\ee 
Operators
\begin{eqnarray}\label{Qbr}
&\mathcal{Q}_1=
\left(
\begin{array}{cc}
 0 & A_b   \\
A^\#_b  &  0
\end{array}
\right),\qquad
\mathcal{Q}_2=i\sigma_3\mathcal{Q}_1\,,&
\end{eqnarray}
are the supercharges of the system 
(\ref{Hextendbr}). They
satisfy the relations
\be
[\mathcal{H},\mathcal{Q}_a]=0,\qquad
\{\mathcal{Q}_a,\mathcal{Q}_b\}=2\delta_{ab}(\mathcal{H}-\delta^{-2})\,.
\ee
The partner Hamiltonians  can also be 
intertwined by the second order 
operators $A^{\alpha_2}_1A^{\alpha_1\#}_1$
and $A^{\alpha_1}_1A^{\alpha_2\#}_1$
via  the Hamiltonian of the intermediate (`virtual' here) free particle 
system\,: $A^{\alpha_2}_1A^{\alpha_1\#}_1H^{\alpha_1}_1=
A^{\alpha_2}_1H_0A^{\alpha_1\#}_1=
H^{\alpha_2}_1 A^{\alpha_2}_1A^{\alpha_1\#}_1$,
$A^{\alpha_1}_1A^{\alpha_2\#}_1H^{\alpha_2}_1=
A^{\alpha_1}_1H_0A^{\alpha_2\#}_1=
H^{\alpha_1}_1A^{\alpha_1}_1A^{\alpha_2\#}_1$.
Hence, system  (\ref{Hextendbr})  is  characterized
additionally
by the second order supercharges
\be\label{Sbr}
\mathcal{S}_1=
\left(
\begin{array}{cc}
 0 &   A^{\alpha_2}_1A^{\alpha_1\#}_1 \\ 
A^{\alpha_1}_1A^{\alpha_2\#}_1  &  0
\end{array}
\right),\qquad
\mathcal{S}_2=i\sigma_3\mathcal{S}_1\,,
\ee
\be
[\mathcal{H},\mathcal{S}_a]=0,\qquad
\{\mathcal{S}_a,\mathcal{S}_b\}=2\delta_{ab}\mathcal{H}^2\,.
\ee
We note here that the first order intertwining operators 
from which supercharges (\ref{Qbr}) are composed can be obtained 
by employing the reduction of intertwining operators mentioned in
Section \ref{SecSusyExact}.
Indeed,
the Hamiltonians $H^{\alpha_1}_1$ and 
$H^{\alpha_2}_1$ can be intertwined by the third 
order differential operator $A^{\alpha_2}_1P_0A^{\alpha_1\#}_1$ by using the chain of
equalities 
 $A^{\alpha_2}_1P_0A^{\alpha_1\#}_1H^{\alpha_1}_1=A^{\alpha_2}P_0H_0A^{\alpha_1\#}_1=
 A^{\alpha_2}_1H_0P_0A^{\alpha_1\#}_1=H^{\alpha_2}_1A^{\alpha_2}_1P_0A^{\alpha_1\#}_1$.
 However, using equation (4.18) from ref. \cite{MatPly},
 or by a direct computation, one finds the relation 
  $A^{\alpha_2}_1P_0A^{\alpha_1\#}_1=H^{\alpha_2}_1A_b-i\delta^{-1}A^{\alpha_2}_1A^{\alpha_1\#}_1$
  that shows that the indicated third order intertwining operator
  reduces to the intertwining operators 
  appearing in the  right-upper corners of  the 
  supercharges $\mathcal{Q}_1$ and $\mathcal{S}_1$ multiplied, respectively, 
  by the second order operator $H^{\alpha_2}_1$ and the constant $-i\delta^{-1}$.

The anti-commutator of the 
first and second order supercharges 
generates the matrix Lax-Novikov integral 
$\mathcal{L}_1=\text{diag}\,(P^{\alpha_2}_1,P^{\alpha_1}_1)$,
\be
\{\mathcal{Q}_a,\mathcal{S}_b\}=2\left(\epsilon_{ab}\mathcal{L}_1+i\delta_{ab}\delta^{-1}\mathcal{H}\right).
\ee
This operator  satisfies relations
$
[\mathcal{L}_1, \mathcal{H}]=[\mathcal{L}_1,\mathcal{Q}_a]=
[\mathcal{L}_1,\mathcal{S}_a]=0$,
and plays the role of the central charge 
of the exotic nonlinear $\mathcal{N}=4$ supersymmetry
generated by $\mathcal{H}$, $\mathcal{Q}_a$,
$\mathcal{S}_a$ and $\mathcal{L}_1$. 
Unlike  (\ref{Hextend1x}), system 
(\ref{Hextendbr}) is in  the phase of 
the partially broken exotic 
nonlinear $\mathcal{N}=4$ supersymmetry. 
Its two zero energy eigenstates 
$\Psi^t_{\alpha_2}=({(x+i\alpha_2)^{-1}},0)^t$
and $\Psi^t_{\alpha_1}=(0, {(x+i\alpha_1)^{-1}})^t$
are zero modes 
of
both second order supercharges 
$\mathcal{S}_a$, but neither of them 
is annihilated by
the
first order supercharges $\mathcal{Q}_a$
\cite{MatPly}.

System (\ref{Hextendbr}) is also characterized by the
dynamical integrals $\mathcal{D}=\text{diag}\,(D^{\alpha_2}_1,D^{\alpha_1}_1)$
and $\mathcal{K}=\text{diag}\,(K^{\alpha_2}_1,K^{\alpha_1}_1)$.
Commuting  the  Lax-Novikov integral  repeatedly with the dynamical integral 
$\mathcal{K}$, we generate three more bosonic dynamical integrals of motion,
\begin{eqnarray}
&[\mathcal{L}_1,\mathcal{K}]=-6i\mathcal{G}_1\,,\qquad
[\mathcal{G}_1,\mathcal{K}]=-4i\mathcal{V}\,,\qquad
[\mathcal{V},\mathcal{K}]=-2i\mathcal{R}\,,\qquad
[\mathcal{R},\mathcal{K}]=0\,,&\\
&\mathcal{G}_1=(\Xi\mathcal{H}+\mathcal{A}^\#)-2t\mathcal{L}_1\,,\qquad
\mathcal{V}=i\Xi^2\mathcal{A}^\#-4t\mathcal{G}_1-4t^2\mathcal{L}_1\,,&\\
&\mathcal{R}=\Xi^3 -6t\mathcal{V}-12t^2\mathcal{G}_1-8t^3\mathcal{L}_1\,,&
\end{eqnarray}
where 
$\Xi=\text{diag}\,(\xi_2,\xi_1)$, 
$\mathcal{A}^\#=\text{diag}\,(A^{\alpha_2\#}_1,A^{\alpha_1\#}_1)$.
All these bosonic integrals have the same scaling dimensions 
as their analogs 
in the system from Section \ref{SecSymH1}.
The peculiarity of the system  (\ref{Hextendbr}), however, is that its supercharges
$\mathcal{Q}_a$ and $\mathcal{S}_a$
are not eigenstates of the dilatation generator
under its adjoint action\,:
\begin{eqnarray}\label{DQSnoneig}
&[\mathcal{D},\mathcal{Q}_a]=\frac{1}{2}\delta \mathcal{S}_a\,,\qquad
[\mathcal{D},\mathcal{S}_a]=\frac{3}{2}i \mathcal{S}_a-\frac{1}{2}\delta \mathcal{H}
\mathcal{Q}_a\,.&
\end{eqnarray}
Coherently with this, the way in which the odd dynamical 
integrals for the system  (\ref{Hextendbr}) are 
generated is also different in comparison with 
that for the system from Section 
\ref{SecSusyExact}.
The commutation of the dynamical integral  $\mathcal{K}$
with the first and second order supercharges $\mathcal{Q}_a$ and $\mathcal{S}_a$ 
generates the unique pair of the  odd dynamical integrals $\mu_a$\,:
\be
[\mathcal{K}, \mathcal{Q}_a]=\delta^2 \mathcal{Q}_a+ 2\delta \mu_a\,,
\qquad
[\mathcal{K}, \mathcal{S}_a]=\delta^2 \mathcal{S}_a +3i \delta \mathcal{Q}_a-4\delta 
\mathcal{D}\mathcal{Q}_a +6i\mu_a \,,
\ee
\be\label{muabr}
\mu_1=i\Xi\sigma_1\mathcal{A}^{\#}
 -2t\mathcal{S}_1\,,\qquad \mu_2=i\sigma_3\mu_1\,.
\ee
Commutation of $\mathcal{K}$ with $\mu_a$
produces a new pair of  fermionic integrals $\Gamma_a$, 
\begin{eqnarray}
&[\mathcal{K}, \mu_a]=4i\Gamma_a+\delta^2 \mu_a -2\delta \mathcal{K} \mathcal{Q}_a\,,&\\
\label{Gamma1def}
&\Gamma_1=(\Xi^2\sigma_1-\delta \Xi\sigma_2) -(4\mu_1 +2\delta Q_1)t-4t^2 S_1\,,
\qquad \Gamma_2=i\sigma_3\Gamma_1\,.&
\end{eqnarray}
The time-independent  term in (\ref{Gamma1def}) is presented equivalently in the form
$\xi_1\xi_2\sigma_1$.
Finally, commutation of $\mathcal{K}$ with $\Gamma_a$ generates
dynamical odd integrals $\Omega_a$,
\begin{eqnarray}
&[\mathcal{K}, \Gamma_a]=\Omega_a\,,&\\
&\Omega_1=\left(3\delta^2\Xi^2\sigma_1+
(2\delta\Xi^3-\delta^3\Xi)\sigma_2\right)
+
t\left(-2\delta^3 \mathcal{Q}_1 -12\delta^2\mu_1  +12\delta \mathcal{K} \mathcal{Q}_1 -12i\Gamma_1\right)&\nonumber\\
&+t^2 \left(-12\delta^2 \mathcal{S}_1 +48\delta\mathcal{D}Q_1 -24i\mu_1 \right)
+t^3
\left(-16i \mathcal{S}_1 +16 \delta\mathcal{H}\mathcal{Q}_1\right)\,,&\label{TildGam}
\end{eqnarray}
$\Omega_2=i\sigma_3\Omega_1$.
The commutation relation
\begin{eqnarray}
&[\mathcal{K}, \Omega_a]=\delta^2\left(4\Omega_a-2\delta^2 {\Gamma}_a-\frac{1}{2}
\mathcal{K}\Gamma_a\right)&
\end{eqnarray}
then signals that the process of generation of odd dynamical integrals terminates.
Commutation relations 
\begin{eqnarray}
&[\mathcal{D},\mu_a]=\frac{i}{2}\mu_a-\delta\left(\mathcal{D}-\frac{i}{4}\right)\mathcal{Q}_a\,,
\qquad
[\mathcal{D},\Gamma_a]=-\frac{i}{2}\Gamma_a-\frac{1}{2}\delta\mathcal{K}\mathcal{Q}_a+
\frac{1}{2}\delta^2\mu_a\,,&\label{DmuGam}\\
&[\mathcal{D},\Omega_a]=-\frac{3}{2}i\Omega_a-\frac{3}{2}\delta^3\mathcal{K}\mathcal{Q}_a+
2i\delta^2 \Gamma_a
+\delta^2\left( \frac{1}{2}\delta^2-\mathcal{K}\right)\mu_a&\label{DmuTilGam}
\end{eqnarray}
show that, like $\mathcal{Q}_a$ and $\mathcal{S}_a$,
neither of the odd dynamical integrals 
is eigenstate under the adjoint action of the dilation generator.
The commutation relations of the odd dynamical integrals
with the Hamiltonian operator are 
\begin{eqnarray}
&[\mathcal{H}, \mu_a]=-2iS_a\,,
\qquad
[\mathcal{H}, \Gamma_a]=-2i(2\mu_a+\delta Q_a)\,,
&\\
&[\mathcal{H}, \Omega_a]=12\Gamma_a 
-i\delta^2(10\mu_a +3\delta Q_a)
+4\delta^2 \mathcal{D}(\mu_a+\delta Q_a)
+12i\mathcal{K} Q_a\,.
&
\end{eqnarray}
The anti-commutation relations of dynamical integrals 
$\mu_a$ with $\mu_b$, 
$\mathcal{Q}_b$ and
$\mathcal{S}_b$ are
\begin{eqnarray}
&\{\mu_a,\mu_b\}=8
\delta_{ab}\left(\mathcal{D}^2-4i\mathcal{D}+2i\delta\mathcal{G}_2-9\mathcal{I}\right)\,,&\\
&\{\mathcal{Q}_a,\mu_b\}=\delta_{ab}\left( -\delta\mathcal{H}+4i\delta^{-1}\mathcal{D}+\delta^{-1}\mathcal{I}\right)+
\epsilon_{ab}\left(2\mathcal{G}_1+i\delta\Sigma\mathcal{H}\right)\,,&\\
&\{\mathcal{S}_a,\mu_b\}=\delta_{ab}
\left(-3i\mathcal{H}+4\mathcal{D}\mathcal{H}+i\delta\mathcal{L}_2\right)
+\epsilon_{ab}\left(-2\Sigma\mathcal{H}+\delta\mathcal{L}_1\right)\,,&
\end{eqnarray}
where $\mathcal{L}_2=\sigma_3\mathcal{L}_1$ and 
$\mathcal{G}_2=\sigma_3\mathcal{G}_1$ are two 
additional  bosonic  integrals.  
Instead of them, one can take their linear combinations
with  $\mathcal{L}_1$ and $\mathcal{G}_1$ to obtain
$\mathcal{L}_-=\Pi_-\mathcal{L}_1$
and $\mathcal{G}_-=\Pi_-\mathcal{G}_1$, which could be considered 
as analogs of  the integrals 
(\ref{mathP-}) of the system (\ref{Hextend1x}).
Finally, the system under consideration is described
by ten bosonic integrals $\mathcal{H}$, 
$\mathcal{D}$, $\mathcal{K}$, $\Sigma$, $\mathcal{L}_1$,
$\mathcal{G}_1$, $\mathcal{V}$, $\mathcal{R}$,
$\mathcal{L}_2$ and $\mathcal{G}_2$
(or by $\mathcal{L}_-$ and $\mathcal{G}_-$ instead of the two last
integrals),
the same number of  fermionic integrals 
$\mathcal{Q}_a$,  $\mathcal{S}_a$, 
$\mu_a$, $\Gamma_a$ and $\Omega_a$, and by 
the central charge $\mathcal{I}$.

Essential peculiarity   here is  
that 
the $PT$-regularization parameter $\alpha$ of the dimension
of length does not appear in the superalgebra of the system  (\ref{Hextend1x}), 
while in the  
(anti)-commutation relations  for the system 
(\ref{Hextendbr})
there appears
the parameter $\delta=\alpha_1-\alpha_2$.

We will not write down the missing  (anti)-commutation relations of the integrals,
but  examine  the relation between 
the systems  (\ref{Hextend1x}) and 
(\ref{Hextendbr}) in the light of  their nonlinearly 
supersymmetrically extended and deformed  conformal symmetries.
First of all, we  observe  that the nonlinear extensions 
of the conformal $sl(2,\R)$ symmetry  generated 
by their sets of the bosonic integrals are very similar.
The only difference appears in commutation relations 
of the corresponding pairs of the operators ($\mathcal{P}_-$, $\mathcal{G}_-$)
and ($\mathcal{L}_-$, $\mathcal{G}_-$). 
More essential difference is that 
 system  (\ref{Hextend1x}) contains the algebra of the superconformal
$osp(2\vert 2)$ symmetry  as a sub-superalgebra, whereas 
 system (\ref{Hextendbr})  does not have such a sub-superalgebra.
 This is reflected, particularly,  in  the peculiarity of the latter system
 encoded in relations (\ref{DQSnoneig}), (\ref{DmuGam}) and 
(\ref{DmuTilGam}), cf.  (\ref{SQDLMK}).
Since in the limit  $\alpha_1\rightarrow \infty$ with identification
$\alpha_2=\alpha$ the superpotential $\mathcal{W}_b(x)$ of 
the system  (\ref{Hextendbr}) transforms into superpotential 
$\mathcal{W}_e(x)$ of the system (\ref{Hextend1x}),
and the Hamiltonian (\ref{Hextendbr}) transforms into 
 (\ref{Hextend1x}), the essential difference between 
 two systems at first glance may seem
to be rather surprising. 
To clarify this point,  let us see what happens in the indicated limit 
with other operators. We have $A_b\rightarrow A^\alpha_1$, 
$A_b^\#\rightarrow A^{\alpha\#}_1$, 
$A^{\alpha_1}_1\rightarrow \frac{d}{dx}$
and
$A^{\alpha_1\#}_1\rightarrow -\frac{d}{dx}$.
As a result, we find that the generators $\mathcal{Q}_a$,
$\mathcal{S}_a$ and $\mathcal{L}_1$ of the system 
(\ref{Hextendbr}) transform into the corresponding integrals 
$\mathcal{Q}_a$,
$\mathcal{S}_a$ and $\mathcal{L}$
of the system (\ref{Hextend1x}),
while  $\mathcal{L}_-$
transforms into the integral 
$\mathcal{P}_-$ multiplied with the 
Hamiltonian (\ref{Hextend1x}).
The limit applied to other integrals is more delicate, however.
The dilatation generator  of the system (\ref{Hextendbr}) 
with $\alpha_2=\alpha$ is represented equivalently in the form
$\mathcal{D}=\mathcal{D}(x+i\alpha) +\frac{i}{2}(\alpha_1-\alpha)\mathcal{P}_-$,
where by $\mathcal{D}(x+i\alpha)$ we indicate  
 the dilatation generator for the system
(\ref{Hextend1x}). So, the integral 
$\mathcal{D}$ transforms into $\mathcal{D}(x+i\alpha)$  if the  limit $\alpha_1\rightarrow\infty $ 
is accompanied by a `renormalization' which consists in omitting
the  integral $\frac{i}{2}\mathcal{P}_-$ multiplied with  the factor 
$\delta=(\alpha_1-\alpha)\rightarrow \infty$.
A similar picture is valid  for relation between 
other bosonic dynamical integrals of both systems 
in the indicated limit.
Particularly, we have $\mathcal{K}=\mathcal{K}(x+i\alpha)
+2i\delta\mathcal{J}_-(x+i\alpha)-\delta^2\Pi_-$, and this integral transforms
into $\mathcal{K}(x+i\alpha)$ by making the same kind of 
 `renormalization'.
 For fermionic dynamical integral $\mu_1$ the relation
 $\mu_1=\mu_1(x+\alpha)-\delta\Pi_-\mathcal{Q}_1$
 is valid when $\alpha_1\rightarrow\infty $, and, as a result, 
 a `renormalized' form of the integrals (\ref{muabr}) will correspond to the
 dynamical integrals (\ref{mu1}) of the system (\ref{Hextend1x}).
For the integral $\Gamma_1$ with $\alpha_1\rightarrow\infty $ 
we have $\Gamma_1=\kappa_1(x+i\alpha)-i\delta\lambda_2(x+i\alpha)$,
and the   `renormalized' form of the integrals
(\ref{Gamma1def}) gives us the integrals (\ref{kappa}) 
for the system  (\ref{Hextend1x}).
This also means  that the limit $\alpha_1\rightarrow\infty$
applied to the rescaled integrals $\delta^{-1}\Gamma_a$ 
reproduces the dynamical integrals 
(\ref{lambda}).
Analogously,  
the appropriately 
rescaled integrals (\ref{TildGam}) in the limit  $\alpha_1\rightarrow\infty$
reproduce the integrals (\ref{lambda})\,:
$\delta^{-3}\Omega_a\rightarrow \lambda_a(x+i\alpha)$.

All this shows that in the limit when one of the two 
parameters $\alpha_1$ or $\alpha_2$ is sent to infinity,
 the system (\ref{Hextendbr}) 
in the phase of the partially broken 
 exotic nonlinear $\mathcal{N}=4$ super-Poincar\'e symmetry
 transforms into the system  (\ref{Hextend1x})  
in  the unbroken phase,
and all the integrals of the latter system can be reproduced
from those of the former system.
The relation between the integrals, however,  is rather non-trivial,
that is behind  the essential difference between the 
nonlinear super-extended versions of the conformal symmetry
generated in two cases.

\section{Discussion and outlook}\label{Outlook}

The AFF model \cite{deAFF} is obtained from the model (\ref{Hg})
by adding into  the Hamiltonian operator the confining harmonic potential term $x^2$
that effectively results in introduction of  another boundary condition $\psi(+\infty)=0$
at another edge of the interval $x\in (0,+\infty)$ in addition to 
the Dirichlet boundary condition $\psi(0)=0$.
This procedure does not  change the symmetry:
the system described by the ``regularized" AFF Hamiltonian has the same conformal
symmetry as the initial system (\ref{Hg}), but it changes radically the  spectrum.
Instead of continuous non-degenerate  spectrum of the system (\ref{Hg}) with $E>0$, the AFF model 
has the equidistant discrete spectrum  which, up to a constant  shift,   coincides 
with the spectrum of the half-harmonic oscillator \cite{CarInzPly}. This is not surprising
as at particular values of  the coupling constant $g=n(n+1)$,
the AFF model is generated from the half-harmonic oscillator by 
the appropriate nonsingular on the half-line 
Darboux-Crum transformation \cite{CarInzPly}.
{}From this point of view the AFF model with confining harmonic potential term
is  rather a rational deformation of the half-harmonic oscillator  than the deformation
(by adding the harmonic term) of the two-particle 
Calogero system (\ref{Hg}) with the omitted center of mass coordinate.
Nevertheless, in this way the ``regularization" recipe accepted in \cite{deAFF} 
solves the problem of the absence of the ground state in the 
initial system (\ref{Hg}).
One could not restrict the values of $x$ to the half-line in the system (\ref{Hg})
as well as in the ``regularized" AFF model and consider both systems
on  the whole real line $x\in \R$. Potential with the inverse squared  term is not
penetrable at $x=0$, and the probability flux between regions $x<0$ and $x>0$ will 
be equal to zero, i.e. in this case we effectively will have  two copies of the system
with doubly degenerate either continuous spectrum with $E>0$ in the case   of (\ref{Hg})
or the discrete spectrum in the AFF model \cite{deAFF}. 
Besides the conformal symmetry, system  (\ref{Hg}) has another very important peculiarity.
As we noted, it's potential  with $g=n(n+1)$ is an algebro-geometric 
solution to the stationary KdV equation
or the higher equation of its hierarchy, which is 
characterized by the existence of the differential operator of order $2n+ 1$, related with a
higher order Novikov equation  \cite{Novikov}. 
This higher order Lax-Novikov operator  commutes with the Hamiltonian operator
(\ref{Hg}) \cite{BelBob}. 
In this picture, a free particle on the whole real line can be treated as a zero-gap ($n=0$) system for which 
 the corresponding first order Lax-Novikov differential operator is just the momentum integral 
 $-i\frac{d}{dx}$. But unlike the free particle case with $x\in\R$,  the  Lax-Novikov 
 differential operator of order $2n+1$ for the quantum system (\ref{Hg}) with $g=n(n+1)$
 is not a true integral of motion since it takes out the wave functions 
 from the domain of the Hamiltonian operator (\ref{Hg}), 
 as it also happens for
 the operator $-i\frac{d}{dx}$ for the free particle on the half-line.  
 The $PT$-regularization  of  the system (\ref{Hg})
 we apply, $x\rightarrow x+i\alpha$, $\R\ni \alpha\neq 0$, $x>0~\rightarrow x\in \R$, 
 allows us to transform the Lax-Novikov operator of the algebro-geometric method of solution 
 of the KdV equation and higher equations of its hierarchy, into the true 
integral for the quantum system $H^\alpha_n=-\frac{d^2}{dx^2}+\frac{n(n+1)}{(x+i\alpha)^2}$,
which turns out to be a Darboux-dressed momentum of the free particle on the whole line.
As a result, the symmetry of the $PT$-regularized system is described 
by nonlinearly  extended generalized Shr\"odinger algebra which includes conformal
$sl(2,\R)$ algebra as the subalgebra, and there appears a non-degenerate bound state 
$\psi_0=1/(x+i\alpha)^n$ of zero energy
at the very edge of its doubly degenerate continuous spectrum,
which is nothing else as the Darboux-Crum transformed zero energy 
free particle's state $\psi_0=1$.
As we showed, such extension of symmetry also has  very profound consequences for 
supersymmetric  $PT$-regularized conformal mechanics systems
in comparison with the superconformal  $osp(2|2)$ symmetry 
of the superextended version of the system (\ref{Hg}).
Particularly, in the $PT$-regularized superconformal mechanics system
in the partially broken phase of the exotic nonlinear $\mathcal{N}=4$ super-Poincar\'e symmetry
we considered in the previous section,  the
$osp(2|2)$  is not contained at all as a sub-superalgebra
in the corresponding nonlinearly super-extended Schr\"odinger  algebra.

\vskip0.1cm
As it was shown in \cite{MatPly}, the  systems we considered here   
are the simplest representatives of a broader class of the systems which 
can be generated by applying  the chains of Darboux transformations  to the  quantum free particle. 
The interesting question is what happens with the conformal symmmetry 
for more complicated systems from the indicated class, particularly,
in the systems  related  to solutions of the  equations of the KdV hierarchy 
which reveal 
the  properties of   the extreme waves \cite{MatPly}.
\vskip0.1cm

Since conformal mechanics (\ref{Hg}) with  special values of the coupling constant $g=n(n+1)$ 
plays a special role  in  the  Huygens' principle \cite{Veselov}, 
the interesting question  that deserves a separate   investigation  corresponds to 
the treatment of the considered systems  as (1+1)-dimensional field 
theories  from the perspective of the associated time-dependent  Schr\"odinger 
equation.  An a priori  complication which may  appear in this  direction 
is that due to the higher-derivative nature 
of the Lax-Novikov operators, it is impossible, at least directly, to associate
them with Noether integrals of the corresponding field systems.
The indicated field-theoretical  generalization is also interesting 
bearing in mind that the considered $PT$-symmetric quantum mechanical systems
are also related to  the singular kinks arising as traveling waves 
in the Liouville and $SU(3)$ conformal Toda systems  \cite{MatPly}.
\vskip0.1cm

The obtained  results could be generalized for  the case of the AFF conformal
mechanics model with the confining 
harmonic potential term. The essential difference in such a case  is  that the corresponding 
systems do not have  Lax-Novikov integrals. However, in the case of
rational deformations of such  systems with  special values of the 
coupling constant $g=n(n+1)$, instead of the Lax-Novikov integrals 
they are characterized  by higher-derivative spectrum-generating 
ladder operators \cite{CarPly,CarInzPly}. As a result, in that case also there appear
nonlinear  extensions of the (super)-Schr\"odinger  symmetry 
of the structure similar to that investigated here \cite{InzPly+}.

It would be interesting to generalize our results for the
case of appropriately $PT$-regularized 
many-particle superconformal mechanics \cite{FreeMen}--\cite{KL}.
There,  additional integrals also can be constructed in the form of higher-order
Lax-Novikov type integrals  via the Darboux-dressing procedure
\cite{CorLecPly}. It is necessary to stress that the formal Lax-Novikov type integrals   
in $n$-particle Calogero-Moser systems with $n>2$ also have  a pure quantum nature,
and they are not related to (maximal) super-integrability of the systems which takes place already at the classical level
\cite{Wojc,Kuznetsov,Gonera,Ranada}. Similarly to the present case of  $n=2$, 
the square of the indicated formal integrals also reduces there to a polynomial in the Liouville
charges \cite{CorLecPly}.
 
\vskip0.3cm

\noindent {\bf Acknowledgements}
\vskip 0.2cm

We acknowledge support from research projects FONDECYT 1130017 and Convenio Marco
Universidades del Estado (Project USA1555), Chile, and MINECO (Project MTM2014-57129-
C2-1-P), Spain. JMG also acknowledges the Junta de Castilla y Le\'on for financial support
under grant VA057U16. MSP is grateful for the warm hospitality at Salamanca 
University where this work started.



\begin{thebibliography}{99}
\bibitem{deAFF}
 V.~de Alfaro, S.~Fubini and G.~Furlan,
  \emph{ ``Conformal invariance in quantum mechanics,''}
  \href{https://link.springer.com/article/10.1007%2FBF02785666}{
  Nuovo Cim.\ A {\bf 34}, 569 (1976)}.

  


   \bibitem{Calogero} 
  F.~Calogero,
   \emph{ ``Solution of the one-dimensional N body problems with quadratic and/or inversely 
   quadratic pair potentials,''}
    \href{http://aip.scitation.org/doi/10.1063/1.1665604}{J.\ Math.\ Phys.\  {\bf 12}, 419 (1971)}.
  
  \bibitem{AkuPash}
   V.~P.~Akulov and A.~I.~Pashnev,
   \emph{ ``Quantum superconformal model in (1,2) space,''}
  \href{https://link.springer.com/article/10.1007%2FBF01086252}{Theor.\ Math.\ Phys.\  {\bf 56}, 862 (1983)}
  [Teor.\ Mat.\ Fiz.\  {\bf 56}, 344 (1983)].

\bibitem{FubRabi}
    S.~Fubini and E.~Rabinovici,
   \emph{ ``Superconformal quantum mechanics,''}
 \href{https://www.sciencedirect.com/science/article/pii/055032138490422X?via%3Dihub}{ Nucl.\ Phys.\ B {\bf 245}, 17 (1984)}.
 


\bibitem{IvaKriLev1} 
  E.~A.~Ivanov, S.~O.~Krivonos and V.~M.~Leviant,
  \emph{``Geometry of conformal mechanics,''}
 \href{http://iopscience.iop.org/article/10.1088/0305-4470/22/4/005/meta}{J.\ Phys.\ A {\bf 22}, 345 (1989)}.
  
  \bibitem{IvaKriLev2} 
  E.~A.~Ivanov, S.~O.~Krivonos and V.~M.~Leviant,
  \emph{ ``Geometric superfield approach to superconformal mechanics,''}
 \href{http://iopscience.iop.org/article/10.1088/0305-4470/22/19/015/meta}{J.\ Phys.\ A {\bf 22}, 4201 (1989)}.

\bibitem{FreeMen} 
  D.~Z.~Freedman and P.~F.~Mende,
  \emph{``An exactly solvable $N$ particle system in supersymmetric quantum mechanics,''}
  \href{https://www.sciencedirect.com/science/article/pii/055032139090364J?via%3Dihub}{Nucl.\ Phys.\ B {\bf 344}, 317 (1990)}.


 \bibitem{Wyll}
  N.~Wyllard,
  \emph{ ``(Super)conformal many body quantum mechanics with extended supersymmetry,''}
  \href{https://aip.scitation.org/doi/10.1063/1.533273}{J.\ Math.\ Phys.\  {\bf 41}, 2826 (2000)}
  \href{https://arxiv.org/abs/hep-th/9910160}{\textcolor{magenta}{[hep-th/9910160]}}.
  

  
  \bibitem{BGIK}
  S.~Bellucci, A.~Galajinsky, E.~Ivanov and S.~Krivonos,
  \emph{``AdS(2)/CFT(1), canonical transformations and superconformal mechanics,''}
  \href{https://www.sciencedirect.com/science/article/pii/S0370269303000406?via%3Dihub}{Phys.\ Lett.\ B {\bf 555}, 99 (2003)}
    \href{https://arxiv.org/abs/hep-th/0212204}{\textcolor{magenta}{[arXiv:hep-th/0212204]}}.
 
  \bibitem{BGK}
   S. Bellucci, A. Galajinsky, S. Krivonos,
 \emph{``New many-body superconformal models as reductions of simple composite systems,"}
\href{https://journals.aps.org/prd/abstract/10.1103/PhysRevD.68.064010}{Phys. Rev. D 68 (2003) 064010}
 \href{https://arxiv.org/abs/hep-th/0304087}{\textcolor{magenta}{[arXiv:hep-th/0304087]}}.
 
   \bibitem{BGL}
  S.~Bellucci, A.~V.~Galajinsky and E.~Latini,
 \emph{``New insight into WDVV equation,''}
  \href{https://journals.aps.org/prd/abstract/10.1103/PhysRevD.71.044023}{Phys.\ Rev.\ D {\bf 71}, 044023 (2005)}
   \href{https://arxiv.org/abs/hep-th/0411232}{\textcolor{magenta}{[arXiv:hep-th/0411232]}}.
  
  
 \bibitem{GLP}
  A.~Galajinsky, O.~Lechtenfeld and K.~Polovnikov,
   \emph{``Calogero models and nonlocal conformal transformations,''}
  \href{https://www.sciencedirect.com/science/article/pii/S0370269306013736?via%3Dihub}{Phys.\ Lett.\ B {\bf 643}, 221 (2006)}
      \href{https://arxiv.org/abs/hep-th/0607215}{\textcolor{magenta}{[arXiv:hep-th/0607215]}};
  \emph{``N=4 superconformal Calogero models,''}
 \href{http://iopscience.iop.org/article/10.1088/1126-6708/2007/11/008/meta}{JHEP {\bf 0711}, 008 (2007)}
     \href{https://arxiv.org/abs/0708.1075}{\textcolor{magenta}{[arXiv:0708.1075 [hep-th]]}}.
  
     \bibitem{KL}
  S.~Krivonos and O.~Lechtenfeld,
   \emph{``Many-particle mechanics with D(2,1;$\alpha$) superconformal symmetry,''}
  \href{https://link.springer.com/article/10.1007%2FJHEP02%282011%29042}{JHEP {\bf 1102}, 042 (2011)}
       \href{https://arxiv.org/abs/1012.4639}{\textcolor{magenta}{[arXiv:1012.4639 [hep-th]]}}.



  
 \bibitem{AdS/CFT1}
  J.~M.~Maldacena,
  \emph{ ``The Large N limit of superconformal field theories and supergravity,''}
  \href{https://link.springer.com/article/10.1023%2FA%3A1026654312961}{Int.\ J.\ Theor.\ Phys.\  {\bf 38}, 1113 (1999)}
  [\href{http://www.intlpress.com/site/pub/pages/journals/items/atmp/content/vols/0002/0002/a001/}{Adv.\ Theor.\ Math.\ Phys.\  {\bf 2}, 231 (1998)}]
 \href{https://arxiv.org/abs/hep-th/9711200}{\textcolor{magenta}{[hep-th/9711200}}.
  
  \bibitem{AdS/CFT2}
  S.~S.~Gubser, I.~R.~Klebanov and A.~M.~Polyakov,
  \emph{``Gauge theory correlators from noncritical string theory,''}
 \href{https://www.sciencedirect.com/science/article/pii/S0370269398003773?via%3Dihub}{ Phys.\ Lett.\ B {\bf 428}, 105 (1998)}
 \href{https://arxiv.org/abs/hep-th/9802109}{\textcolor{magenta}{[hep-th/9802109]}}.
  
  \bibitem{AdS/CFT3}
  E. Witten, \emph{``Anti-de Sitter space and holography,''}  \href{http://www.intlpress.com/site/pub/pages/journals/items/atmp/content/vols/0002/0002/a002/}{Adv. 
  Theor.  Math.  Phys.  {\bf 2}, 253 (1998)} 
  \href{https://arxiv.org/abs/hep-th/9802150}{\textcolor{magenta}{[hep-th/9802150]}}.



\bibitem{CDKKTV} 
  P.~Claus, M.~Derix, R.~Kallosh, J.~Kumar, P.~K.~Townsend and A.~Van Proeyen,
  \emph{ ``Black holes and superconformal mechanics,''}
  \href{https://journals.aps.org/prl/abstract/10.1103/PhysRevLett.81.4553}{Phys.\ Rev.\ Lett.\  {\bf 81}, 4553 (1998)}
   \href{https://arxiv.org/abs/hep-th/9804177}{\textcolor{magenta}{[hep-th/9804177]}}.

\bibitem{AIPT} 
  J.~A.~de Azcarraga, J.~M.~Izquierdo, J.~C.~Perez Bueno and P.~K.~Townsend,
 \emph{ ``Superconformal mechanics and nonlinear realizations,''}
 \href{https://journals.aps.org/prd/abstract/10.1103/PhysRevD.59.084015}{Phys.\ Rev.\ D {\bf 59}, 084015 (1999)}
   \href{https://arxiv.org/abs/hep-th/9810230}{\textcolor{magenta}{[hep-th/9810230]}}.


\bibitem{GibTow} 
  G.~W.~Gibbons and P.~K.~Townsend,
  \emph{ ``Black holes and Calogero models,''}
  \href{https://www.sciencedirect.com/science/article/pii/S037026939900266X?via%3Dihub}{Phys.\ Lett.\ B {\bf 454}, 187 (1999)}
 \href{https://arxiv.org/abs/hep-th/9812034}{\textcolor{magenta}{[hep-th/9812034]}}.


\bibitem{MichStro} 
  J.~Michelson and A.~Strominger,
  \emph{``Superconformal multiblack hole quantum mechanics,''}
  \href{http://iopscience.iop.org/article/10.1088/1126-6708/1999/09/005/meta}{JHEP {\bf 9909}, 005 (1999)}
   \href{https://arxiv.org/abs/hep-th/9908044}{\textcolor{magenta}{[hep-th/9908044]}}.

   \bibitem{deTerDosBro} 
  G.~F.~de Teramond, H.~G.~Dosch and S.~J.~Brodsky,
  \emph{``Baryon spectrum from superconformal quantum mechanics and its light-front 
  holographic embedding,''}
 \href{https://journals.aps.org/prd/abstract/10.1103/PhysRevD.91.045040}{ Phys.\ Rev.\ D {\bf 91}, no. 4, 045040 (2015)}
   \href{https://arxiv.org/abs/1411.5243}{\textcolor{magenta}{  [arXiv:1411.5243 [hep-ph]]}}.
  
  
   \bibitem{BTDL} 
  S.~J.~Brodsky, G.~F.~de Tramond, H.~G.~Dosch and C.~Lorc,
  \emph{``Universal effective hadron dynamics from superconformal algebra,''}
  \href{https://www.sciencedirect.com/science/article/pii/S0370269316302155?via%3Dihub}{Phys.\ Lett.\ B {\bf 759}, 171 (2016)}
  \href{https://arxiv.org/abs/1604.06746}{\textcolor{magenta}{  [arXiv:1604.06746 [hep-ph]]}}.



\bibitem{DHVin} 
  E.~D'Hoker and L.~Vinet,
  \emph{``Dynamical supersymmetry of the magnetic monopole and the $1/r^2$ potential,''}
 \href{https://link.springer.com/article/10.1007%2FBF01213405}{Commun.\ Math.\ Phys.\  {\bf 97}, 391 (1985).}


  
\bibitem{MichStro+} 
  J.~Michelson and A.~Strominger,
  \emph{``The geometry of (super)conformal quantum mechanics,''}
 \href{https://link.springer.com/article/10.1007%2FPL00005528}{ Commun.\ Math.\ Phys.\  {\bf 213}, 1 (2000)}
    \href{https://arxiv.org/abs/hep-th/9907191}{\textcolor{magenta}{[hep-th/9907191]}}.


    \bibitem{CacKleZan} 
  S.~Cacciatori, D.~Klemm and D.~Zanon,
 \emph{ ``W(infinity) algebras, conformal mechanics, and black holes,''}
  \href{http://iopscience.iop.org/article/10.1088/0264-9381/17/8/301/meta}{Class.\ Quant.\ Grav.\  {\bf 17}, 1731 (2000)}
   \href{https://arxiv.org/abs/hep-th/9910065}{\textcolor{magenta}{[hep-th/9910065]}}.

  

  
  \bibitem{BPMSV} 
  R.~Britto-Pacumio, J.~Michelson, A.~Strominger and A.~Volovich,
  \emph{``Lectures on superconformal quantum mechanics and multi-black hole moduli spaces,''}
  \href{https://link.springer.com/chapter/10.1007%2F978-94-011-4303-5_6}{NATO Sci.\ Ser.\ C {\bf 556}, 255 (2000)}
   \href{https://arxiv.org/abs/hep-th/9911066}{\textcolor{magenta}{[hep-th/9911066]}}.
  
  
 
  
  \bibitem{Papado} 
  G.~Papadopoulos,
  \emph{``Conformal and superconformal mechanics,''}
  \href{http://iopscience.iop.org/article/10.1088/0264-9381/17/18/310/meta}{Class.\ Quant.\ Grav.\  {\bf 17}, 3715 (2000)}
   \href{https://arxiv.org/abs/hep-th/0002007}{\textcolor{magenta}{[hep-th/0002007]}}.
  
  
  
  \bibitem{DonPasRivTsu} 
  E.~E.~Donets, A.~Pashnev, V.~O.~Rivelles, D.~P.~Sorokin and M.~Tsulaia,
  \emph{``N=4 superconformal mechanics and the potential structure of AdS spaces,''}
  \href{https://www.sciencedirect.com/science/article/pii/S0370269300006705?via%3Dihub}{Phys.\ Lett.\ B {\bf 484}, 337 (2000)}
  \href{https://arxiv.org/abs/hep-th/0004019}{\textcolor{magenta}{[hep-th/0004019]}}.
  
  \bibitem{Plyushchay:2000hb} 
  M.~S.~Plyushchay,
  \emph{``Monopole Chern-Simons term: Charge-monopole system as a particle with spin,''}
\href{https://www.sciencedirect.com/science/article/pii/S0550321300005307?via%3Dihub}{Nucl.\ Phys.\ B {\bf 589}, 413 (2000)}
  \href{https://arxiv.org/abs/hep-th/0004032}{\textcolor{magenta}{[hep-th/0004032]}}.
  
  
     \bibitem{Gho} 
  P.~K.~Ghosh,
  \emph{ ``Conformal symmetry and the nonlinear Schrodinger equation,''}
  \href{https://journals.aps.org/pra/abstract/10.1103/PhysRevA.65.012103}{Phys.\ Rev.\ A {\bf 65}, 012103 (2002)}
     \href{https://arxiv.org/abs/cond-mat/0102488}{\textcolor{magenta}{[cond-mat/0102488]}}.
    
  \bibitem{GunKoeNic} 
  M.~Gunaydin, K.~Koepsell and H.~Nicolai,
    \emph{``The minimal unitary representation of E(8(8)),''}
  \href{http://www.intlpress.com/site/pub/pages/journals/items/atmp/content/vols/0005/0005/a003/}{Adv.\ 
  Theor.\ Math.\ Phys.\  {\bf 5}, 923 (2002)}
  \href{https://arxiv.org/abs/hep-th/0109005}{\textcolor{magenta}{[hep-th/0109005]}}.
  
 
    
    
  \bibitem{PiolWal} 
  B.~Pioline and A.~Waldron,
  \emph{``Quantum cosmology and conformal invariance,''}
  \href{https://journals.aps.org/prl/abstract/10.1103/PhysRevLett.90.031302}{Phys.\ Rev.\ Lett.\  {\bf 90}, 031302 (2003)}
     \href{https://arxiv.org/abs/hep-th/0209044}{\textcolor{magenta}{[hep-th/0209044]}}.
  
    \bibitem{CamOrd} 
  H.~E.~Camblong and C.~R.~Ordonez,
  \emph{``Anomaly in conformal quantum mechanics: From molecular physics to black holes,''}
 \href{https://journals.aps.org/prd/abstract/10.1103/PhysRevD.68.125013}{Phys.\ Rev.\ D {\bf 68}, 125013 (2003)}
     \href{https://arxiv.org/abs/hep-th/0303166}{\textcolor{magenta}{[hep-th/0303166]}}.
  
  \bibitem{LeiPly1} 
  C.~Leiva and M.~S.~Plyushchay,
 \emph{ ``Conformal symmetry of relativistic and nonrelativistic systems and AdS/CFT correspondence,''}
  \href{https://www.sciencedirect.com/science/article/pii/S0003491603001180?via%3Dihub}{Annals Phys.\  {\bf 307}, 372 (2003)}
 \href{https://arxiv.org/abs/hep-th/0301244}{\textcolor{magenta}{[hep-th/0301244]}}.
  

  
  
  \bibitem{PioWal} 
  B.~Pioline and A.~Waldron,
  \emph{``Automorphic forms: A Physicist's survey,''} In:  
  \href{https://link.springer.com/chapter/10.1007%2F978-3-540-30308-4_7}{
  Frontiers in Number Theory, Physics, and Geometry II,
277 (2007), Springer}
     \href{https://arxiv.org/abs/hep-th/0312068}{\textcolor{magenta}{[hep-th/0312068]}}.
  
  \bibitem{BelGalLat}
  S.~Bellucci, A.~V.~Galajinsky and E.~Latini,
  \emph{``New insight into the Witten-Dijkgraff-Verlinde-Verlinde equation,''}
  \href{https://journals.aps.org/prd/abstract/10.1103/PhysRevD.71.044023}{Phys.\ Rev.\ D {\bf 71}, 044023 (2005)}
   \href{https://arxiv.org/abs/hep-th/0411232}{\textcolor{magenta}{[hep-th/0411232]}}.
  
    \bibitem{GaiStroYin} 
  D.~Gaiotto, A.~Strominger and X.~Yin,
 \emph{ ``Superconformal black hole quantum mechanics,''}
  \href{http://iopscience.iop.org/article/10.1088/1126-6708/2005/11/017/meta}{JHEP {\bf 0511}, 017 (2005)}
   \href{https://arxiv.org/abs/hep-th/0412322}{\textcolor{magenta}{[hep-th/0412322]}}.
  
  \bibitem{MelSam} 
  S.~Meljanac and A.~Samsarov,
  \emph{``Universal properties of conformal quantum many-body systems,''}
  \href{https://www.sciencedirect.com/science/article/pii/S0370269305004004?via%3Dihub}{Phys.\ Lett.\ B {\bf 613}, 221 (2005)}
  Erratum: \href{https://www.sciencedirect.com/science/article/pii/S0370269305008567?via%3Dihub}{[Phys.\ Lett.\ B {\bf 620}, 200 (2005)]}
   \href{https://arxiv.org/abs/hep-th/0503174}{\textcolor{magenta}{[hep-th/0503174]}}.
    
 
  
  \bibitem{DuvGibHorj} 
  C.~Duval, G.~W.~Gibbons and P.~Horvathy,
 \emph{ ``Celestial mechanics, conformal structures and gravitational waves,''}
 \href{https://journals.aps.org/prd/abstract/10.1103/PhysRevD.43.3907}{ Phys.\ Rev.\ D {\bf 43}, 3907 (1991)}
      \href{https://arxiv.org/abs/hep-th/0512188}{\textcolor{magenta}{[hep-th/0512188]}}.
  
  

\bibitem{AnaGomZan} 
  A.~Anabalon, J.~Gomis, K.~Kamimura and J.~Zanelli,
  \emph{``N=4 superconformal mechanics as a non linear realization,''}
  \href{http://iopscience.iop.org/article/10.1088/1126-6708/2006/10/068/meta}{JHEP {\bf 0610}, 068 (2006)}
  \href{https://arxiv.org/abs/hep-th/0607124}{\textcolor{magenta}{[hep-th/0607124]}}.
 
   \bibitem{Gunaydin:2007bg} 
  M.~Gunaydin, A.~Neitzke, B.~Pioline and A.~Waldron,
   \emph{ ``Quantum attractor flows,''}
  \href{http://iopscience.iop.org/article/10.1088/1126-6708/2007/09/056/meta}{JHEP {\bf 0709}, 056 (2007)}
     \href{https://arxiv.org/abs/0707.0267}{\textcolor{magenta}{  [arXiv:0707.0267 [hep-th]]}}.

  
    \bibitem{HakNer} 
  T.~Hakobyan and A.~Nersessian,
  \emph{``Lobachevsky geometry of (super)conformal mechanics,''}
 \href{https://www.sciencedirect.com/science/article/pii/S0375960109000930?via%3Dihub}{ Phys.\ Lett.\ A {\bf 373}, 1001 (2009)}
    \href{https://arxiv.org/abs/0803.1293}{\textcolor{magenta}{[arXiv:0803.1293 [hep-th]]}}.
  
      \bibitem{MalMarTac} 
  J.~Maldacena, D.~Martelli and Y.~Tachikawa,
  \emph{``Comments on string theory backgrounds with non-relativistic conformal symmetry,''}
  \href{http://iopscience.iop.org/article/10.1088/1126-6708/2008/10/072/meta}{JHEP {\bf 0810}, 072 (2008)}
     \href{https://arxiv.org/abs/0807.1100}{\textcolor{magenta}{[arXiv:0807.1100 [hep-th]]}}.
  
  
  \bibitem{CorJakPly} 
  F.~Correa, V.~Jakubsky and M.~S.~Plyushchay,
  \emph{``Aharonov-Bohm effect on AdS(2) and nonlinear supersymmetry of reflectionless P\"oschl-Teller system,''}
  \href{https://www.sciencedirect.com/science/article/pii/S000349160900044X?via%3Dihub}{Annals Phys.\  {\bf 324}, 1078 (2009)}
   \href{https://arxiv.org/abs/0809.2854}{\textcolor{magenta}{  [arXiv:0809.2854 [hep-th]]}}.
  

  
  
  \bibitem{AlvCorHorPly} 
  P.~D.~Alvarez, J.~L.~Cortes, P.~A.~Horvathy and M.~S.~Plyushchay,
  \emph{``Super-extended noncommutative Landau problem and conformal symmetry,''}
  \href{http://iopscience.iop.org/article/10.1088/1126-6708/2009/03/034/meta}{JHEP {\bf 0903}, 034 (2009)}
  \href{https://arxiv.org/abs/0901.1021}{\textcolor{magenta}{[arXiv:0901.1021 [hep-th]]}}.




  
  \bibitem{BagGop} 
  A.~Bagchi and R.~Gopakumar,
  \emph{``Galilean conformal algebras and AdS/CFT,''}
  \href{http://iopscience.iop.org/article/10.1088/1126-6708/2009/07/037/meta}{JHEP {\bf 0907}, 037 (2009)}
  \href{https://arxiv.org/abs/0902.1385}{\textcolor{magenta}{ [arXiv:0902.1385 [hep-th]]}}.

 
  \bibitem{CorFalJakPly} 
  F.~Correa, H.~Falomir, V.~Jakubsky and M.~S.~Plyushchay,
  \emph{``Hidden superconformal symmetry of spinless Aharonov-Bohm system,''}
  \href{http://iopscience.iop.org/article/10.1088/1751-8113/43/7/075202/meta}{J.\ Phys.\ A {\bf 43}, 075202 (2010)}
   \href{https://arxiv.org/abs/0906.4055}{\textcolor{magenta}{[arXiv:0906.4055 [hep-th]]}}.



  \bibitem{HakKruLecNer} 
  T.~Hakobyan, S.~Krivonos, O.~Lechtenfeld and A.~Nersessian,
  \emph{``Hidden symmetries of integrable conformal mechanical systems,''}
  \href{https://www.sciencedirect.com/science/article/pii/S0375960109015254?via%3Dihub}{Phys.\ Lett.\ A {\bf 374}, 801 (2010)}
    \href{https://arxiv.org/abs/0908.3290}{\textcolor{magenta}{ [arXiv:0908.3290 [hep-th]]}}.

  
  \bibitem{ChaKacPiSan} 
  C.~Chamon, R.~Jackiw, S.~Y.~Pi and L.~Santos,
  \emph{``Conformal quantum mechanics as the CFT$_1$ dual to AdS$_2$,''}
 \href{https://www.sciencedirect.com/science/article/pii/S0370269311006447?via%3Dihub}{ Phys.\ Lett.\ B {\bf 701}, 503 (2011)}
   \href{https://arxiv.org/abs/1106.0726}{\textcolor{magenta}{  [arXiv:1106.0726 [hep-th]]}}.
 
   
  \bibitem{KuzTop} 
  Z.~Kuznetsova and F.~Toppan,
  \emph{``D-module representations of N=2,4,8 superconformal algebras and their superconformal mechanics,''}
  \href{https://aip.scitation.org/doi/10.1063/1.4705270}{J.\ Math.\ Phys.\  {\bf 53}, 043513 (2012)}
  \href{https://arxiv.org/abs/1112.0995}{\textcolor{magenta}{[arXiv:1112.0995 [hep-th]]}}.
  
  
  \bibitem{FedIvaLec} 
  S.~Fedoruk, E.~Ivanov and O.~Lechtenfeld,
  \emph{ ``Superconformal mechanics,''}
  \href{http://iopscience.iop.org/article/10.1088/1751-8113/45/17/173001/meta}{J. Phys. A {\bf 45}, 173001 (2012)}
   \href{https://arxiv.org/abs/1112.1947}{\textcolor{magenta}{[arXiv:1112.1947 [hep-th]]}}.
  
  \bibitem{AndGonMas} 
  K.~Andrzejewski, J.~Gonera and P.~Maslanka,
  \emph{``Nonrelativistic conformal groups and their dynamical realizations,''}
 \href{https://journals.aps.org/prd/abstract/10.1103/PhysRevD.86.065009}{ Phys.\ Rev.\ D {\bf 86}, 065009 (2012)}
    \href{https://arxiv.org/abs/1204.5950}{\textcolor{magenta}{ [arXiv:1204.5950 [math-ph]]}}.
 
 

\bibitem{Gonera} 
  J. Gonera, \emph{ ``Conformal mechanics,''}
\href{https://www.sciencedirect.com/science/article/pii/S0003491613000912?via%3Dihub}{ Annals Phys. {\bf 335}, 61 (2013)}
  \href{https://arxiv.org/abs/1211.4403}{\textcolor{magenta}{[arXiv:1211.4403 [hep-th]]}}.
 
 
\bibitem{MolVil} 
  J.~Molina-Vilaplana and G.~Sierra,
  \emph{``An $xp$ model on $AdS_2$ spacetime,''}
  \href{https://www.sciencedirect.com/science/article/pii/S0550321313004847?via%3Dihub}{Nucl.\ Phys.\ B {\bf 877}, 107 (2013)}
    \href{https://arxiv.org/abs/1212.2436}{\textcolor{magenta}{[arXiv:1212.2436 [hep-th]]}}.
 
 
   \bibitem{PlyWip} 
  M.~S.~Plyushchay and A.~Wipf,
  \emph{``Particle in a self-dual dyon background: hidden free nature, and exotic superconformal symmetry,''}
  \href{https://journals.aps.org/prd/abstract/10.1103/PhysRevD.89.045017}{Phys.\ Rev.\ D {\bf 89}, no. 4, 045017 (2014)}
   \href{https://arxiv.org/abs/1311.2195}{\textcolor{magenta}{[arXiv:1311.2195 [hep-th]]}}.

  \bibitem{Brod} 
  S.~J.~Brodsky, G.~F.~de Teramond, H.~G.~Dosch and J.~Erlich,
  \emph{``Light-front holographic QCD and emerging confinement,''}
  \href{https://www.sciencedirect.com/science/article/pii/S0370157315002306?via%3Dihub}{Phys.\ Rept.\  {\bf 584}, 1 (2015)}
     \href{https://arxiv.org/abs/1407.8131}{\textcolor{magenta}{  [arXiv:1407.8131 [hep-ph]]}}.
  


  


\bibitem{Andrz} 
  K.~Andrzejewski, J.~Gonera, P.~Kosinski and P.~Maslanka,
 \emph{ ``On dynamical realizations of $l$-conformal Galilei groups,''}
 \href{https://www.sciencedirect.com/science/article/pii/S0550321313004069?via%3Dihub}{ Nucl.\ Phys.\ B {\bf 876}, 309 (2013)}
     \href{https://arxiv.org/abs/1305.6805}{\textcolor{magenta}{[arXiv:1305.6805 [hep-th]]}}.
  

  
  \bibitem{MasRod} 
  M.~Masuku and J.~P.~Rodrigues,
  \emph{``De Alfaro, Fubini and Furlan from multi matrix systems,''}
  \href{https://link.springer.com/article/10.1007%2FJHEP12%282015%29175}{JHEP {\bf 1512}, 175 (2015)}
     \href{https://arxiv.org/abs/1509.06719}{\textcolor{magenta}{[arXiv:1509.06719 [hep-th]]}}.
  

  

  
  \bibitem{Master} 
  I.~Masterov,
  \emph{``Remark on higher-derivative mechanics with l-conformal Galilei symmetry,''}
 \href{https://aip.scitation.org/doi/10.1063/1.4963169}{ J.\ Math.\ Phys.\  {\bf 57}, no. 9, 092901 (2016)}
    \href{https://arxiv.org/abs/1607.02693}{\textcolor{magenta}{[arXiv:1607.02693 [hep-th]]}}.




\bibitem{Andr} 
  K.~Andrzejewski,
  \emph{``Quantum conformal mechanics emerging from unitary representations of SL(2,$\mathbb{R}$),''}
 \href{https://www.sciencedirect.com/science/article/pii/S0003491616000312?via%3Dihub}{ Annals Phys.\  {\bf 367}, 227 (2016)}
   \href{https://arxiv.org/abs/1506.05596}{\textcolor{magenta}{[arXiv:1506.05596 [hep-th]]}}.

\bibitem{BonCorLatWal} 
  R.~Bonezzi, O.~Corradini, E.~Latini and A.~Waldron,
  \emph{``Quantum mechanics and hidden superconformal symmetry,''}
  \href{https://journals.aps.org/prd/abstract/10.1103/PhysRevD.96.126005}{Phys.\ Rev.\ D {\bf 96}, no. 12, 126005 (2017)}
   \href{https://arxiv.org/abs/1709.10135}{\textcolor{magenta}{[arXiv:1709.10135 [hep-th]]}}.
  



\bibitem{MatPly} 
  J.~Mateos Guilarte and M.~S.~Plyushchay,
  \emph{``Perfectly invisible $\mathcal{PT}$-symmetric zero-gap systems, 
  conformal field theoretical kinks, and exotic nonlinear supersymmetry,''}
 \href{https://link.springer.com/article/10.1007%2FJHEP12%282017%29061}{ JHEP {\bf 1712}, 061 (2017)}
   \href{https://arxiv.org/abs/1710.00356}{\textcolor{magenta}{[arXiv:1710.00356 [hep-th]]}}.
  
  
  
  \bibitem{InzPly} 
  L.~Inzunza and M.~S.~Plyushchay,
  \emph{``Hidden superconformal symmetry: Where does it come from?,''}
  \href{https://journals.aps.org/prd/abstract/10.1103/PhysRevD.97.045002}{Phys.\ Rev.\ D {\bf 97}, no. 4, 045002 (2018)}
   \href{https://arxiv.org/abs/1711.00616}{\textcolor{magenta}{[arXiv:1711.00616 [hep-th]]}}.


 \bibitem{AirMcKMos}
H. Airault, H. P. McKean, and J. Moser,
  \emph{``Rational and elliptic solutions of the 
  Korteweg-de Vries equation and a related many-body problem,"}
  \href{http://onlinelibrary.wiley.com/doi/10.1002/cpa.3160300106/abstract}{Comm. Pure Appl. Math. {\bf 30}, 95 (1977)}.

\bibitem{AdlMos}
M. Adler and J. Moser, 
 \emph{ ``On a Class of Polynomials
Connected with the Korteweg-deVries Equation,"}
\href{https://link.springer.com/article/10.1007/BF01609465}{Commun. Math. Phys. {\bf 61}, 1 (1978)}.


 
   \bibitem{GorNek} 
  A.~Gorsky and N.~Nekrasov,
   \emph{``Hamiltonian systems of Calogero type and two-dimensional Yang-Mills theory,''}
  \href{http://www.sciencedirect.com/science/article/pii/0550321394904294?via%3Dihub}{Nucl.\ Phys.\ B {\bf 414}, 213 (1994)}
\href{https://arxiv.org/abs/hep-th/9304047}{\textcolor{magenta}{[hep-th/9304047]}}.

\bibitem{DuiGru}
J. J. Duistermaat and F. A. Gr\"unbaum,
   \emph{``Differential equations in the spectral parameter,"}
   \href{https://link.springer.com/article/10.1007/BF01206937}{Comm. Math. Phys. {\bf 103}, 177 (1986)}.
   

\bibitem{Veselov}
A. P. Veselov, \emph{``Huygens' principle and integrability,"}
\href{https://link.springer.com/chapter/10.1007/978-3-0348-8898-1_17}{Prog. in Math., {\bf 169}, 259 (1998)}.
  
 \bibitem{NovZak} 
 S. P. Novikov, S.V. Manakov, L. P. Pitaevskii, and V. E. Zakharov, \textsl{Theory of Solitons} (Plenum,
New York, 1984).

\bibitem{Krich}
I. M. Krichever, 
 \emph{``Baker-Akhiezer functions and integrable systems,"}
in: ``Integrability. The Seiberg-Witten and Whitham Equations", Edited by H. W. Braden
and I. M. Krichever, p. 1 (Gordon and Breach Science Publishers, Amsterdam, 2000).

\bibitem{BelBob}
E. D. Belokolos, A. I. Bobenko, V. Z. EnolÕskii, A. R. Its, V. B. Matveev,  \textsl{Algebro-Geometric
Approach to Nonlinear Integrable Equations}  (Springer, Berlin, 1994).

\bibitem{CPLax} 
 F.~Correa and M.~S.~Plyushchay,
  ``Hidden supersymmetry in quantum bosonic systems,''
  \href{https://www.sciencedirect.com/science/article/pii/S0003491606002831?via%3Dihub}{Annals Phys.\  {\bf 322}, 2493 (2007)}
    \href{https://arxiv.org/abs/hep-th/0605104}{\textcolor{magenta}{[arXiv:hep-th/0605104]}}.
  
  \bibitem{CNPLax} 
  F.~Correa, L.~M.~Nieto and M.~S.~Plyushchay,
 \emph{``Hidden nonlinear supersymmetry of finite-gap Lame equation,''}
  \href{https://www.sciencedirect.com/science/article/abs/pii/S0370269306014274?via%3Dihub}{Phys.\ Lett.\ B {\bf 644}, 94 (2007)}
   \href{https://arxiv.org/abs/hep-th/0608096}{\textcolor{magenta}{[arXiv:hep-th/0608096]}}.

\bibitem{BC}
J.L. Burchnall, T.W. Chaundy,
 \emph{``Commutative ordinary differential operators,"}
\href{https://londmathsoc.onlinelibrary.wiley.com/doi/abs/10.1112/plms/s2-21.1.420}{
Proc. London Math. Soc. {\bf s2-21}, 420 (1923)};
\href{https://royalsocietypublishing.org/doi/abs/10.1098/rspa.1928.0069}{Proc. Royal Soc. London A {\bf 118}, 557  (1928)}.

\bibitem{Ince}
E. L. Ince, 
\textsl{Ordinary differential equations} (Dover, 1956).

\bibitem{CorJakNP} 
  F.~Correa, V.~Jakubsky, L.~M.~Nieto and M.~S.~Plyushchay,
    \emph{``Self-isospectrality, special supersymmetry, and their effect on the band structure,''}
  \href{https://journals.aps.org/prl/abstract/10.1103/PhysRevLett.101.030403}{Phys.\ Rev.\ Lett.\  {\bf 101}, 030403 (2008)}
    \href{https://arxiv.org/abs/0801.1671}{\textcolor{magenta}{[arXiv:0801.1671 [hep-th]]}}.

\bibitem{AraMatPly} 
  A.~Arancibia, J.~Mateos Guilarte and M.~S.~Plyushchay,
  \emph{ ``Effect of scalings and translations on the supersymmetric quantum mechanical structure of soliton systems,''}
  \href{https://journals.aps.org/prd/abstract/10.1103/PhysRevD.87.045009}{Phys.\ Rev.\ D {\bf 87}, no. 4, 045009 (2013)}
    \href{https://arxiv.org/abs/1210.3666}{\textcolor{magenta}{  [arXiv:1210.3666 [math-ph]]}}.

\bibitem{LeiPly} 
  C.~Leiva and M.~S.~Plyushchay,
   \emph{ ``Superconformal mechanics and nonlinear supersymmetry,''}
 \href{http://iopscience.iop.org/article/10.1088/1126-6708/2003/10/069/meta}{ JHEP {\bf 0310}, 069 (2003)}
    \href{https://arxiv.org/abs/hep-th/0304257}{\textcolor{magenta}{   [hep-th/0304257]}}.
  
  

\bibitem{CorOlPly} 
  F.~Correa, M.~A.~del Olmo and M.~S.~Plyushchay,
  \emph{ ``On hidden broken nonlinear superconformal symmetry of conformal mechanics and nature of double nonlinear 
  superconformal symmetry,''}
  \href{https://www.sciencedirect.com/science/article/pii/S0370269305013821?via%3Dihub}{Phys.\ Lett.\ B {\bf 628}, 157 (2005)}
   \href{https://arxiv.org/abs/hep-th/0508223}{\textcolor{magenta}{  [hep-th/0508223]}}.



  \bibitem{BenRev} 
  C. M. Bender,  \emph{``Making sense of non-Hermitian Hamiltonians,"}
 \href{http://iopscience.iop.org/article/10.1088/0034-4885/70/6/R03/meta}{Rept. Prog. Phys. {\bf 70}, 947 (2007)}
   \href{https://arxiv.org/abs/hep-th/0703096}{\textcolor{magenta}{[arXiv:hep-th/0703096]}}.  
 
  \bibitem{Most}  
 A. Mostafazadeh, 
  \emph{ ``Pseudo-Hermitian Representation of Quantum Mechanics,"}
\href{https://www.worldscientific.com/doi/abs/10.1142/S0219887810004816}{Int. J. Geom. Meth. Mod. Phys. {\bf 7}, 1191 (2010)}
   \href{https://arxiv.org/abs/0810.5643}{\textcolor{magenta}{[arXiv:0810.5643 [quant-ph]]}}.  


\bibitem{Wigner}
E. P. Wigner, 
 \emph{``Normal form of antiunitary operators,"}
 \href{https://aip.scitation.org/doi/abs/10.1063/1.1703672?journalCode=jmp}{J. Math. Phys. {\bf 1}, 409 (1960)}.

\bibitem{Defects1} 
  A.~Arancibia, F.~Correa, V.~Jakubsk\'y, J.~Mateos Guilarte and M.~S.~Plyushchay,
  \emph{ ``Soliton defects in one-gap periodic system and exotic supersymmetry,''}
  \href{https://journals.aps.org/prd/abstract/10.1103/PhysRevD.90.125041}{Phys.\ Rev.\ D {\bf 90}, no. 12, 125041 (2014)}
  \href{https://arxiv.org/abs/1410.3565}{\textcolor{magenta}{[arXiv:1410.3565 [hep-th]]}}.

\bibitem{Defects2} 
  A.~Arancibia and M.~S.~Plyushchay,
    \emph{``Chiral asymmetry in propagation of soliton defects in crystalline backgrounds,''}
 \href{https://journals.aps.org/prd/abstract/10.1103/PhysRevD.92.105009}{Phys.\ Rev.\ D {\bf 92}, no. 10, 105009 (2015)}
   \href{https://arxiv.org/abs/1507.07060}{\textcolor{magenta}{[arXiv:1507.07060 [hep-th]]}}.


\bibitem{BenBoe}
C.M. Bender, S. Boettcher, 
    \emph{``Real spectra in non-hermitian Hamiltonians having $\mathcal{PT}$ symmetry,''}
 \href{https://journals.aps.org/prl/abstract/10.1103/PhysRevLett.80.5243}{Phys. Rev. Lett. {\bf 80}, 5243 (1998)}  
    \href{https://arxiv.org/abs/physics/9712001}{\textcolor{magenta}{[arXiv:physics/9712001]}}.
 
 

 \bibitem{DDT1}
P. Dorey, C. Dunning and R. Tateo,  \emph{``Spectral equivalences, Bethe Ansatz equations,
and reality properties in $\mathcal{PT}$-symmetric quantum mechanics,"}
\href{http://iopscience.iop.org/article/10.1088/0305-4470/34/28/305/meta}{J. Phys. A 34, 5679 (2001)}
 \href{https://arxiv.org/abs/hep-th/0103051}{\textcolor{magenta}{[arXiv:hep-th/0103051]}}. 

 
 

 \bibitem{DDT2}
  P.~Dorey, C.~Dunning and R.~Tateo,
 \emph{``Supersymmetry and the spontaneous breakdown of $\mathcal{PT}$ symmetry,''}
  \href{http://iopscience.iop.org/article/10.1088/0305-4470/34/28/102/meta}{J.\ Phys.\ A {\bf 34}, L391 (2001)}
  \href{https://arxiv.org/abs/hep-th/0104119}{\textcolor{magenta}{[hep-th/0104119]}}.

  
  
  \bibitem{Znoj}   
  M. Znojil,
   \emph{ ``$\mathcal{PT}$-symmetric harmonic oscillators,"}
  \href{https://www.sciencedirect.com/science/article/pii/S0375960199004296?via}{Phys. Lett. A  {\bf 259},  220 (1999)} 
  \href{https://arxiv.org/abs/quant-ph/9905020}{\textcolor{magenta}{[quant-ph/9905020]}}.
  
  \bibitem{CorFri}  
  F.~Correa and A.~Fring,
  \emph{``Regularized degenerate multi-solitons,''}
 {JHEP {\bf 1609}, 008 (2016)}
   \href{}{\textcolor{magenta}{[arXiv:1605.06371 [nlin.SI]]}}.
  
 \bibitem{FriZno} 
A. Fring and  M. Znojil, 
\emph{``$\mathcal{PT}$-symmetric deformations of Calogero models,"}
\href{http://iopscience.iop.org/article/10.1088/1751-8113/41/19/194010/meta}{J. Phys. A {\bf 41}, 194010 (2008)}
\href{https://arxiv.org/abs/0802.0624}{\textcolor{magenta}{[arXive:0802.0624 [quant-ph]]}}.


 
  \bibitem{CorLech}   
  F. Correa and O. Lechtenfeld, 
  \emph{``$\mathcal{PT}$ deformation of angular Calogero models,"} 
 { JHEP {\bf 1711}, 122 (2017)}
  Ê
  \href{https://arxiv.org/abs/1705.05425}{\textcolor{magenta}{[arXiv:1705.05425 [hep-th]]}}.

  
   
  \bibitem{NatPhys}  
R. El-Ganainy, K. G. Makris, M. Khajavikhan, Z. H. Musslimani, S. Rotter, and  D. N. Christodoulides,
\emph{ ``Non-Hermitian physics and PT symmetry,"}
\href{https://www.nature.com/articles/nphys4323}{Nature Physics {\bf 14}, 11 (2018)}. 

  
\bibitem{CarPly} 
  J.~F.~Cari\~nena and M.~S.~Plyushchay,
 \emph{``ABC of ladder operators for rationally extended quantum harmonic oscillator systems,''}
 \href{http://iopscience.iop.org/article/10.1088/1751-8121/aa739b/meta}{ J.\ Phys.\ A {\bf 50}, no. 27, 275202 (2017)}
    \href{https://arxiv.org/abs/1701.08657}{\textcolor{magenta}{  [arXiv:1701.08657 [math-ph]]}}.


\bibitem{CarInzPly} 
 J.~F.~Cari\~nena, L.~Inzunza and M.~S.~Plyushchay,
   \emph{``Rational deformations of conformal mechanics,''}
   \href{https://journals.aps.org/prd/abstract/10.1103/PhysRevD.98.026017}{Phys.\ Rev.\ D {\bf 98}, no. 2, 026017 (2018)} 
    \href{https://arxiv.org/abs/1707.07357}{\textcolor{magenta}{[arXiv:1707.07357 [math-ph]]}}.

\bibitem{ArPlcrys}
  A.~Arancibia and M.~S.~Plyushchay,
    \emph{``Extended supersymmetry of the self-isospectral crystalline and soliton chains,''}
  \href{https://journals.aps.org/prd/abstract/10.1103/PhysRevD.85.045018}{Phys.\ Rev.\ D {\bf 85}, 045018 (2012)}
     \href{https://arxiv.org/abs/1111.0600}{\textcolor{magenta}{[rXiv:1111.0600 [hep-th]]}}.


\bibitem{Novikov}
S. P. Novikov, 
\emph{``The periodic problem for the Korteweg-de Vries equation,''}
\href{https://link.springer.com/article/10.1007%2FBF01075697}{Funct. Anal. Appl. {\bf 8}, 236 (1974)}.


\bibitem{InzPly+}
 L.~Inzunza and M.~S.~Plyushchay,
  \emph{``Hidden symmetries of rationally deformed superconformal mechanics,"} 
  \href{https://journals.aps.org/prd/abstract/10.1103/PhysRevD.99.025001}{Phys. Rev. D {\bf 99}, 025001(2019)}
      \href{https://arxiv.org/abs/1809.08527}{\textcolor{magenta}{[arXiv:1809.08527 [hep-th]]}}.


\bibitem{CorLecPly} 
 F.~Correa, O.~Lechtenfeld and M.~Plyushchay,
  \emph{``Nonlinear supersymmetry in the quantum Calogero model,''}
  \href{https://link.springer.com/article/10.1007%2FJHEP04%282014%29151}{JHEP {\bf 1404}, 151 (2014)}
   \href{https://arxiv.org/abs/1312.5749}{\textcolor{magenta}{[arXiv:1312.5749 [hep-th]]}}.
 

\bibitem{Wojc}
S. Wojciechowski,
\emph{``Superintegrability of the Calogero-Moser system,"}
\href{https://www.sciencedirect.com/science/article/abs/pii/037596018390018X}{Phys. Lett. A {\bf 95}, 279 (1983)}.

\bibitem{Kuznetsov}
 V.~B.~Kuznetsov,
   \emph{``Hidden symmetry of the quantum Calogero-Moser system,''}
  \href{https://www.sciencedirect.com/science/article/abs/pii/0375960196004215?via%3Dihub}{Phys.\ Lett.\ A {\bf 218}, 212 (1996)}
   \href{https://arxiv.org/abs/solv-int/9509001}{\textcolor{magenta}{[arXiv:solv-int/9509001]}}.

\bibitem{Gonera}
C. Gonera,
    \emph{``A note on superintegrability of the quantum Calogero model,"}
     \href{https://www.sciencedirect.com/science/article/abs/pii/S0375960198009037}{Phys. Lett.  A {\bf 237}, 365 (1998)}.

\bibitem{Ranada}
M. F. Ra\~nada, 
 \emph{``Superintegrability of the Calogero-Moser system: 
 Constants of motion, master symmetries, and time-dependent symmetries,"}
\href{https://aip.scitation.org/doi/10.1063/1.532770}{J. Math. Phys. {\bf 40}, 236 (1999)}.


\end{thebibliography}
\end{document}